\documentstyle[onecolumn,epsfig]{mn}
\newif\ifAMStwofonts

\newcommand{\etal}{{et~al.}}

\newcommand{\ica}{\sc{FastICA}}
\newcommand{\lsim}{\,\lower2truept\hbox{${<\atop\hbox{\raise4truept\hbox{$\sim$}}}$}\,}
\newcommand{\gsim}{\,\lower2truept\hbox{${>\atop\hbox{\raise4truept\hbox{$\sim$}}}$}\,}

\title[All-sky astrophysical component separation with \ica]
{All-sky astrophysical component separation with Fast Independent
Component Analysis (\ica)}

\author[Maino et al.]{D.~Maino$^1\footnote{E-mail: maino@ts.astro.it}$, A.~Farusi$^2$, C.~Baccigalupi$^3$, F.~Perrotta$^4$,
A.J.~Banday$^5$, L.~Bedini$^2$,
\and
C.~Burigana$^6$, G.~De Zotti$^4$, K.M.~G\'orski$^{7,8}$, E. Salerno$^2$\\
$^1$ Osservatorio Astronomico di Trieste, Via G.B. Tiepolo, 11, I-34131 Trieste, Italy\\
$^2$  IEI-CNR, Via Alfieri 1, I-56010 Ghezzano, Pisa, Italy \\
$^3$ SISSA/ISAS, Astrophysics Sector, Via Beirut, 4, I-34014 Trieste, Italy \\
$^4$ Osservatorio Astronomico di Padova, Vicolo dell' Osservatorio 5, I-35122 Padova, Italy \\
$^5$ MPA, Max Planck Inst. f\"ur Astrophysik, Karl-Schwarzschild Str. 1, D-85740 Garching, Germany \\
$^6$ ITeSRE-CNR, Via Gobetti, 101, I-40129 Bologna, Italy \\
$^7$ ESO, European Southern Observatory, Karl-Schwarzschild Str. 2, D-85740 Garching, 
Germany;\\ 
$^8$ Warsaw University Observatory, Aleje Ujazdowskie 4, 00-478 Warszava, Poland}

\pagerange{\pageref{firstpage}--\pageref{lastpage}}
\pubyear{2001}

\begin{document}

\maketitle

\label{firstpage}
\footnotetext{E-mail: maino@ts.astro.it}
\begin{abstract}
We present a new, fast, algorithm for the separation of
astrophysical components superposed in maps of the sky. The
algorithm, based on the Independent Component Analysis (ICA)
technique, is aimed at recovering both the spatial pattern and the
frequency scalings of the emissions from statistically independent
astrophysical processes, present along the line-of-sight, from
multi-frequency observations, without any a priori assumption on
properties of the components to be separated, except that all of
them, but at most one, must have non-Gaussian distributions.

The analysis starts from very simple toy-models of the sky
emission in order to assess the quality of the reconstruction when
inputs are well known and controlled. In particular we study the
dependence of the results of separation conducted on and off the
Galactic plane independently, showing that optimal separation is
achieved for sky regions where components are smoothly
distributed.

Then we move to more realistic applications on simulated
observations of the microwave sky with angular resolution and
instrumental noise at the mean nominal levels for the {\sc Planck}
satellite. We consider several {\sc Planck} observation channels
containing the most important known diffuse signals: the Cosmic
Microwave Background (CMB), Galactic synchrotron, dust and
free-free emissions. A method for calibrating the reconstructed
maps of each component at each frequency has been devised. The
spatial pattern of all the components have been recovered on all
scales probed by the instrument. In particular, the CMB angular
power spectra is recovered at the percent level up to
$\ell_{max}\simeq 2000$.

Frequency scalings and normalization have been recovered with
better than 1\% precision for all the components at frequencies
and in sky regions where their signal-to-noise ratio $\gsim 1.5$;
the error increases at $\sim 10\%$ level for signal-to-noise
ratios $\simeq 1$.

Runs have been performed on a Pentium III 600 MHz computer;
although the computing time slightly depends on the number of
channels and components to be recovered, {\ica} typically took
about 10 minutes for all-sky simulations with 3.5$'$ pixel size.

Although the quoted results have been obtained under a number of
simplifying assumptions, we conclude that {\ica} is an extremely
promising technique for analyzing the maps that will be obtained
by the forthcoming high resolution CMB experiments.
\end{abstract}


\section{Introduction}
\label{introduction}

NASA's Microwave Anisotropy Probe (MAP) satellite (Wright 1999)
has begun its mission aimed at all-sky imaging of the last
scattering surface of CMB photons, at several frequencies and with
high angular resolution. In 2007, the very early phase of our
Universe will be probed with even higher sensitivity and angular
resolution by ESA's {\sc Planck} satellite (Mandolesi et al. 1998;
Puget et al. 1998). In the meantime, new ground-based and
balloon-borne experiments will provide improved maps of regions of
the sky (see de Bernardis et al. 2001 and references therein). The
main goal of these projects is to extract a map as accurate as
possible of CMB anisotropies at all scales relevant for cosmology.
This will allow us to determine the fundamental cosmological
parameters with very high accuracy and to accurately test the
models of structure formation.

Maps produced by these missions will contain not only the CMB
signal but also contributions from different components placed
between us and the last scattering surface. These consist of
Galactic free-free, thermal and spinning dust, and synchrotron
emissions, of compact galactic and extragalactic sources as well
as of the thermal and kinematic Sunyaev-Zeldovich effect in
cluster of galaxies. Therefore, in order to exploit the wealth of
cosmological and astrophysical information encoded in these maps,
it is necessary to separate the different components with high
accuracy and reliability.

A great deal of work in this direction has been carried out by
several authors (de Oliveira-Costa \& Tegmark~1999, Tenorio et
al.~1999, Hobson et al.~1998, 1999, Stolyarov et al.~2001, Prunet
et al.~2001 and Baccigalupi et al.~2000) exploring different
reconstruction techniques, from Wiener filtering and Maximum
Entropy Method (MEM), both in real and Fourier space, to wavelets
and neural networks.

Traditional methods (Wiener filtering and MEM) have been
successfully applied to all-sky maps. They need the prior
knowledge of the frequency dependence of the astrophysical
components to be recovered, as well as of the power spectra of
their spatial pattern. Recently Stolyarov et al.~(2001) have
achieved a high quality reconstruction of the various components
using a MEM algorithm working in harmonic space, assuming prior
knowledge of the frequency dependence and of the
azimuthally-averaged power spectrum of each input component except
for the CMB. Computational requirements are quite demanding: the
reconstruction requires 6 hours using 30 R10000 processors on an
SGI Origin 2000 supercomputer and 14 Gb of RAM.

The Independent Component Analysis (ICA), based on a feed-forward
neural network (Baccigalupi et al.~2000), works remarkably well
without any a-priori knowledge on the components, provided that
they are statistically independent and have all, but at most one,
non-Gaussian distributions, as is actually the case since, at
most, the distribution of CMB fluctuations is Gaussian, while
foregrounds are all highly non-Gaussian. The main advantage of
this approach is that the algorithm is able to learn how to
reconstruct independent components and their frequency scalings
from the input maps. This technique is very interesting for at
least three reasons. First, as we show in this work, it is able to
recover independent components and their frequency dependencies
with high accuracy also in the presence of instrumental noise.
Second, it can be used as prior estimator for non-blind separation
algorithms. Third, a blind approach is necessary whenever reliable
priors for the foreground emissions are not available, as in the
case of CMB polarization.
Baccigalupi et al. (2000) have applied this technique to small sky
patches of $15^\circ\times 15^\circ$, without instrumental noise.
In these ideal conditions, CMB is reconstructed with $1\%$
accuracy, weaker galactic components are recovered with $\sim
10\%$ error, while point sources are extracted down to a flux
corresponding to $\simeq 0.7 \sigma_{\rm CMB}$.

We present here the application of a new ICA method ({\ica}),
proposed by Hyv\"arinen et al.~(1998), to the problem of
separating the various astrophysical components present in all-sky
maps. This technique proved to be extremely fast: on a Pentium III 600 MHz PC,
it took only $\sim 10$ minutes to perform the component separation
over the whole sky with $\sim 3'5$ resolution. Furthermore, and
importantly,  this technique allows us to deal also with the
instrumental noise, at least as long as it can be approximated by
a Gaussian distribution and is uniformly distributed over the sky.


This paper is organized as follows. In Section~\ref{formalization}
we formalize the problem of component separation and list the
assumptions we have made. Section~\ref{fastica} describes in
detail the {\ica} technique used in this work. In
Section~\ref{model} we study the quality of signal reconstruction
with {\ica} for different kinds of toy models i.e. when the input
signals are well known and controlled. Section~\ref{planck} deals
with more realistic simulations of sky maps, focusing on the {\sc
Planck } satellite frequency channels and instrumental
characteristics (noise and FWHM). Finally, a discussion and
conclusions are presented in Section~\ref{conclusions}.
\section{Formalization of the source separation problem}
\label{formalization}
Let us consider the sky radiation as a function of direction
$\hat{r}$ and frequency $\nu$, and assume that it results from the
superposition of signals coming from $N$ different physical
processes.

\begin{equation}
\label{received}
    \tilde{x}(\hat{r} , \nu ) = \sum_{j=1}^{N}\tilde{s}_{j}(\hat{r}, \nu)
\end{equation}
where $\tilde{x}$ is the total radiation from direction $\hat{r}$
at frequency $\nu$, and $\tilde{s}_{j}$ are the individual
physical processes that form the total radiation.

The signal $\tilde{x}$ is observed through a telescope, whose beam
pattern can be modelled, at each frequency, as a shift-invariant
point spread function $B(\hat{r},\nu )$. For each value of $\nu$,
the telescope de-focuses the physical radiation map by convolving
it with the kernel $B$. The frequency-dependent convolved
signal is input to an $M$-channel measuring instrument, which
integrates the signal over frequency on each of its channels and
adds some noise to its output. The final signal at the output of
the measurement channel at frequency $\nu$ is given by

\begin{equation}
x_\nu(\hat{r}) = \int B(\hat{r}-\hat{r}',\nu')
\sum_{j=1}^{N} t_\nu(\nu') \tilde{s}_{j}(\hat{r}',\nu') dr' d\nu' +
\epsilon_\nu(\hat{r})
\label{ichannel}
\end{equation}
where $t_\nu(\nu')$ is the frequency response of the channel, and
$\epsilon_{\nu}(\hat{r})$ is the related noise map. To simplify
the data model in Eq.~(\ref{ichannel}), let us now make the
following assumptions:

{\it Assumption 1}

\noindent Each source signal is a separable function of location
and frequency:

\begin{equation}
\label{ass1}
    \tilde{s}_{j}(\hat{r} , \nu ) = \bar{s}_{j}(\hat{r})
    f_{j}( \nu ) \hspace{3mm}j=1,\ldots,N\
\end{equation}

{\it Assumption 2}

\noindent $B(\hat{r},\nu )$ is frequency-independent over each
channel pass-band i.e. for each frequency where $t_i(\nu)\neq 0$
in the $i$-th channel, the corresponding beam function $B_{i}$ is
the same.
%
%
These two assumptions lead us to a new data model:
\begin{eqnarray}
\label{ich_appr}
x_\nu(\hat{r}) & = & \sum_{j=1}^{N} B_\nu(\hat{r}) \ast
\bar{s}_j(\hat{r}) \cdot \int t_\nu(\nu ') f_j(\nu ')d\nu ' +
\epsilon_\nu(\hat{r}) \nonumber \\
    & = & B_\nu(\hat{r}) \ast \sum_{j=1}^{N} a_{\nu j}
    \bar{s}_j(\hat{r}) + \epsilon_\nu(\hat{r}) \\
\nonumber
\end{eqnarray}
where $*$ denotes convolution, and
\begin{equation}
\label{aij}
     a_{\nu j} = \int_{}^{} t_{\nu}(\nu ') f_{j}(\nu ') d \nu ' .
\end{equation}
%
%
This data model can be further simplified by making
another assumption:

{\it Assumption 3}

\noindent The radiation pattern of the telescope is the same for
all the measurement channels, that is, the function
$B(\hat{r},\nu )$ is frequency independent:
\begin{equation}
\label{radpattern}
B(\hat{r},\nu ) = B(\hat{r})\ .
\end{equation}
In this case, Eq.~(\ref{ich_appr}) can
be rewritten in vector form as:
\begin{equation}
\label{datavector2}
{\bf{x}}(\hat{r}) = {\bf{A}} {\bar{\bf{s}}}(\hat{r}) * B(\hat{r}) +
{\bmath{\epsilon}}(\hat{r})
= {\bf{A}} {\bf{s}}(\hat{r})+ {\bmath{\epsilon}}(\hat{r})
\end{equation}
where each component, $s_{j}$, of the vector ${\bf{s}}$ is given
by the corresponding source function $\bar{s}_{j}$, as defined in
Eq.~(\ref{ass1}), convolved with the frequency-independent
radiation pattern $B$. The matrix ${\bf{A}}$ is the mixing matrix
whose elements are the $a_{\nu j}$ defined by Eq.~(\ref{aij}).

We detailed this derivation to stress all the simplifying
assumptions made. Depending on the particular astrophysical
application, one or more of these assumptions might be not
justified. Each of these cases needs to be addressed by a specific
strategy. In the present simple case, we are just assuming that
the noise is additive, signal-independent, white, Gaussian, and
stationary. For example, it can be noted that, in principle,
assumption 3 is somewhat bad for applications where the instrument
has different $B_{\nu}$ at different frequencies, since it
requires that higher resolution channels are degraded to the worst
one.

Equation (\ref{datavector2}) establishes a point-wise linear data
model for our problem, that is, it holds true independently for
each direction $\hat{r}$ or, likewise, for each pixel in the
vector data map ${\bf{x}}$. At this point, our problem can be
formulated as follows: with reference to the data model
[Eq.~(\ref{datavector2})], for each pixel $p$ identified by a unit
vector $\hat{r}$, estimate both the $N$-dimensional source vector
${\bf{s}}$ and the $M \times N$ mixing matrix $\bf{A}$ from the
$M$-dimensional data vector ${\bf{x}}$.

\section{The {\ica} algorithm}
\label{fastica}

The problem of estimating both ${\bf{A}}$ and ${\bf{s}}$ from
${\bf{x}}$ via Eq.~(\ref{datavector2}) is unsolvable in the
absence of additional information. The ICA approach assumes that

\begin{itemize}
\item the components of vector ${\bf{s}}$ are
independent random processes on the map domains;
\item the components of vector ${\bf{s}}$, save at most one, have
non-Gaussian distributions.
\end{itemize}
In this section, we summarize an efficient strategy, described in
detail in Hyv\"arinen \& Oja (1997) and Hyv\"arinen (1999), to
separate the independent components ${\bf{s}}$, exploiting these
assumptions.

It has been proved (Comon 1994; Cardoso 1998) that if the
components of ${\bf{s}}$ [Eq.~(\ref{datavector2})] are
independent, a solution to our problem can be obtained by
searching a transformation ${\bf{W}}$ for the data vector
${\bf{x}}$ such that the transformed vector ${\bf{y}} = {\bf{W}}
{\bf{x}}$ has independent components. We should thus find a
measure of independence among the components of ${\bf{y}}$ and
maximize it for ${\bf{W}}$. Intuitively, an equivalent approach to
achieve independence is to maximize non-Gaussianity. Indeed, due
to the central limit theorem, a variable resulting from a mixture
of independent variables is ``more Gaussian'' than the original
variables themselves. This means that finding a transformation
such that the Gaussianity of the variables is reduced is
equivalent to find a set of transformed variables that are
``more independent'' than the original ones. A strategy to find
maximally non-Gaussian transformed variables is to select a
suitable measure of non-gaussianity and then maximize it for the
transform operator ${\bf{W}}$. This approach can be shown to be
equivalent to other theoretically sound approaches proposed in the
literature (Comon 1994; Amari \& Chichocki 1998; Cardoso 1998).
Since Eq.~(\ref{datavector2}) is a noisy model, it is necessary to
find a measure of non-Gaussianity, via a suitable defined
neg-entropy, that allows us to adopt computing strategies that are
robust against noise.

An approximation to the neg-entropy function has been proposed by
Hyv\"arinen \& Oja (2000) and Hyv\"arinen (1999) as a measure of
non-Gaussianity for the transformed variables and, if the noise
has the features assumed in the previous section and its
covariance matrix is known, the Gaussian moments of ${\bf{y}}$ are
shown to be robust estimates of the desired function. It is
possible to maximize non-Gaussianity by means of a Newton
algorithm, after some preprocessing of the data. The rows of the
transformation matrix ${\bf{W}}$ can be estimated one by one, and
a noise-robust estimate is achievable. This does not mean that by
linearly transforming ${\bf{x}}$ through ${\bf{W}}$ we obtain a
robust estimate of ${\bf{s}}$, especially in very noisy cases. 
To see this, with no loss of generality, we can assume $M = N$, and
${\bf{W}}$ as a robust estimate of ${\bf{A}}^{-1}$. In this case,
by applying ${\bf{W}}$ to our data, we obtain
\begin{equation}
\label{noisytransform}
{\bf{y}} = {\bf{W}}({\bf{A}} {\bf{s}} + {\bmath{\epsilon }}) =
{\bf{s}} + {\bf{A}}^{-1} {\bmath \epsilon }\ .
\end{equation}
The noise term ${\bf{A}}^{-1} {\bmath{\epsilon} }$ can strongly affect the
high-spatial-frequency components of this solution, and thus some
regularization strategy must normally be adopted to separate the
source signals once ${\bf{W}}$ has been estimated. We intentionally do not
face this point in the present work, which is entirely dedicated to the study
of the capabilities of {\ica} by itself.

Another general feature that one should keep in mind is that the
quality of the estimated matrix ${\bf{W}}$ becomes worse when one
or more signals are much stronger then the others, even if the
reconstructed spatial patterns of the independent components are 
still good. Indeed, suppose for simplicity to have no noise, two
frequency scalings and two independent components only. Then, from
Eq.~(\ref{noisytransform}) it can be easily seen that at each
output $y_{i}$ one has 
\begin{eqnarray}
y_{i}=(W_{i1}A_{11}+W_{i2}A_{21})s_{1}+(W_{i1}A_{12}+W_{i2}A_{22})s_{2}\ .
\label{ratio}
\end{eqnarray}
If, for example, $s_{1}\gg s_{2}$, then a good {\it spatial}
reconstruction for which $y_{i}$ is a copy of $s_{1}$ can be
achieved even in the case where the coefficients of $s_{1}$ and
$s_{2}$ are comparable, or, in other words, the matrix ${\bf{W}}$
is not a robust estimator of ${\bf{A}}^{-1}$.

The algorithm needs a preprocessing step (Hyv\"arinen 1999),
where the data maps are ``quasi whitened''. This step aims at reducing
the number of unknowns with respect to the original problem.
Let us suppose we know the covariance matrix ${\bmath{ \Sigma}} $ of
the system noise. The mean of the data at each frequency is removed
(the offset of each independent component can be recovered in the end
of the separation process, as we show below),
and the covariance matrix ${\bf{C}}$ of the zero mean data is calculated
by just evaluating the following expectation over the set of all available pixels:
\begin{equation}
\label{sigcov}
{\bf{C}} = E\{{\bf{xx}}^T\}\ .
\end{equation}
The preprocessing procedure consists in evaluating a modified noise
covariance and a quasi-whitened data set, respectively, as follows:
\begin{equation}
\label{qwhit1}
\hat{\bmath{\Sigma}} = ({\bf{C}} - {\bmath{\Sigma}})^{-1/2}
{\bmath{\Sigma}} ({\bf{C}} - {\bmath{\Sigma}} )^{-1/2}\ ,
\end{equation}
\begin{equation}
\label{qwhit2}
\hat{\bf{x}} = ({\bf{C}} - {\bmath{\Sigma}})^{-1/2} {\bf{x}}\ .
\end{equation}
The separation algorithm estimates the matrix ${\bf{W}}$ row by
row, thus permitting to recover the components one by one. Let
${\bf w}$ be an $M$-dimensional vector whose dot-product by
$\hat{\bf{x}}$ gives a component of the transformed vector
${\bf{y}} = {\bf{W}} \hat{\bf{x}}$ (${\bf{w}}^T$ is a row of
matrix ${\bf{W}}$). In order to find a noise-robust estimation of
one such vector, we apply the following iterative algorithm,
equipped with a suitable convergence criterion, where $g$ is a
regular non-quadratic function whose form can be chosen as
$g(u)=u^{3}$, $g(u)=\tanh u$, $g(u)=\exp (-u^{2})$ for most
problems (see Hyv\"arinen 1999), and $g'$ is its first derivative:

\begin{enumerate}
\item Choose an initial vector ${\bf w}$;
\item update it through
$$
{\bf{w}}_{new} = E\{\hat{\bf{x}} g({\bf{w}}^T \hat{\bf{x}})\} -
(I+\hat{\bmath\Sigma}){\bf{w}} E\{\ g'({\bf{w}}^T \hat{\bf{x}})\}
$$
where again $E$ denotes expectation over all the available
samples;

\item let
$$
{\bf{w}}_{new} = \frac{{\bf{w}}_{new}}{\|{\bf{w}}_{new}\|}\ .
$$
\item Compare ${\bf{w}}_{new}$ with the old one;
if not converged, go back to (ii), if converged, begin another process.
\end{enumerate}
This procedure maximizes the non-Gaussianity of the component
${\bf{w}}^T\hat{\bf{x}}$. Let us assume that a certain number,
$k$, of rows of $W$ have been evaluated. To estimate the
($k+1$)-th row, we must search it in the subspace orthogonal to
the first $k$ rows. To achieve this goal, an orthogonalization
step ({\it e.g.} Gram-Schmidt) must be inserted between steps (ii)
and (iii).

Once the separation matrix ${\bf{W}}$ has been found, physical
normalization, frequency scalings and offsets of each independent
component in the data can in principle be reconstructed. Consider
normalization and frequency scalings first: by simply inverting
Eq.~(\ref{noisytransform}), we get
\begin{equation}
\label{scalings}
{\bf{x}}={\bf{W}}^{-1}{\bf{y}}\ ,
\end{equation}
which means that the independent component $x_{\nu}^{(j)}$ present
in ${\bf{x}}$ contributes to the signal at $\nu$th frequency:
\begin{equation}
\label{normalization}
x_{\nu}^{(j)}\ ({\rm physical\ units\ at\ frequency}\ \nu)=(W^{-1})_{\nu j}y_{j}\ .
\end{equation}
Then, the frequency scaling for the component $j$ between
frequencies $\nu$ and $\nu'$ is given by
\begin{equation}
\label{scalingsjk}
{x_{\nu j}\over x_{\nu'j}}=
{(W^{-1})_{\nu j}\over (W^{-1})_{\nu'j}}\ .
\end{equation}
The offsets can be recovered as follows. True input data can be
written as the sum of zero mean component plus mean values,
indicated as ${\bf{x}}$ and $\langle{\bf{x}}\rangle$ respectively.
The zero mean component is reconstructed as in
Eq.~(\ref{scalings}), while the total signal obeys:
\begin{equation}
\label{xmean}
\langle{\bf{x}}\rangle+{\bf{x}}=
{\bf{W}}^{-1}({\bf{y}}+\langle{\bf{y}}\rangle)\ .
\end{equation}
Then, by simply taking the mean of this expression we get
\begin{equation}
\label{offsets}
\langle{\bf{y}}\rangle={\bf{W}}\langle{\bf{x}}\rangle\ ,
\end{equation}
so that, at $\nu$th frequency, the offset of the independent
component $j$ is
\begin{equation}
\label{offsets}
\langle x\rangle _{\nu}^{j}\ ({\rm physical\ units\ at\ frequency}\ \nu) =
(W^{-1})_{\nu j}\langle y \rangle _{j}\ .
\end{equation}
Furthermore we can derive the signal-to-noise ratio in the
reconstructed map. Since the system noise covariance matrix
$\bmath{\Sigma}$ is assumed to be known, it is possible to build
noise constrained realizations ${\bf{n}_{x}}$ for each frequency
channel. Once the matrix ${\bf{W}}$ has been recovered, the
corresponding noise realizations in the {\ica} outputs
${\bf{n}_{y}}$ are
\begin{equation}
\label{noise}
{\bf{n}_{y}}={\bf{W}}{\bf{n}_{x}}\ .
\end{equation}
The noise is then transformed like signals and an estimation of
the noise into the reconstructed components is achievable. From
these ``processed'' noise it is straightforward to obtain the
signal-to-noise ratio. This is a useful tool when, e.g., the
number of underlying components is not known a-priori. Suppose, 
for example, that the number $M$ of observing channels is larger
than the number $N$ of the underlying sky components. Using all
the channels leads to the recovery of $M$ ``components'', the
fictitious ones being characterized by a signal-to-noise ratio of
the order, or below, unity.

\section{Toy sky models}
\label{model}

In the next Section we shall show {\ica} results on simulated
skies which are not too far from what the {\sc Planck} satellite
should observe, including the instrumental noise. However, we
preferred to approach that case gradually, inserting here a
Section which starts with highly idealized skies and ends just one
step before the {\sc Planck} case. We want to test the
capabilities of the algorithm when components are well known and
controlled, allowing a better understanding of the more realistic
simulation of the next Section.

\subsection{Toy Model I}
\label{toy1}
\begin{table*}
\begin{center}
\caption{Figures of merit for the reconstruction of the CMB,
convolved with a $7^\circ$ FWHM beam, and of a highly non-Gaussian
signal ($\chi^2$) for different values of $N_{\rm side}$. The
columns ``moments'' report the larger relative error on skewness
and kurtosis. Simulations with (without) $^*$ have the $\chi^2$
signal 3 (30) times stronger than the CMB. Residual are evaluated
for the CMB signal only.} \label{CMBchi2-7deg}
\begin{tabular}{lccccccc}
\hline
\ $N_{\rm side}$ & \multicolumn{2}{c}{$r_S$[\%]} &\multicolumn{2}{c}{$\nu$ scaling[\%]} & \multicolumn{2}{c}{moments[\%]} &
    residuals\\
                 & CMB & $\chi^2$                & CMB & $\chi^2$                       & CMB & $\chi^2$  & \\
\hline
64 & $\gsim 99.99$ & $99.79$ & $\sim 3$ & $\sim 10^{-4}$ &$\sim 0.01$ & $\sim 0.02$ & $\sim 5\times 10^{-4}$ \\
64$^*$&$''$ & $''$ & $\sim 0.35$ & $\sim 10^{-3}$ & $''$  &$''$ & $''$ \\
128& $''$ & $99.88$ & $\sim 2$ & $\sim 10^{-3}$ & $\sim 0.08$& $\sim 0.01$ & $\sim5\times 10^{-3}$\\
128$^*$&$''$& $''$ & $\sim 0.29$ & $\sim 10^{-2}$ & $''$ & $''$& $''$ \\ \hline
\end{tabular}
\end{center}
\end{table*}

In order to evaluate the ability of the {\ica} algorithm to
separate independent sky components, we start with a very simple
toy model, comprising only two components ($N=2$) with different
distributions without, for the moment, taking into account the
instrumental noise. This toy sky model is then ``observed'' at two
frequencies, 53 and 90 GHz; therefore $N=M$. {\ica} is a pointwise
algorithm and we want to verify the dependence of the goodness of
the reconstruction upon the total number of pixels in the maps,
signal statistics and spatial distribution of the signal. We use
sky maps pixelised according to the HEALPix scheme (G\'orski
\etal~1999) and we start working with $N_{\rm side}=64$
corresponding to a pixel size of about $1^\circ$. Our sky emission
model is composed of $(i)$ a standard Cold Dark Matter CMB sky
normalized to $COBE$ (CDM, flat geometry, $\Omega_{b}=0.05$,
$\Omega_{CDM}=0.95$, $h=0.50$, three massless neutrino families,
scale-invariant Gaussian perturbations), which is a realization of
a Gaussian process, smoothed with a $7^\circ$ Gaussian beam, and
$(ii)$ a randomly distributed component (i.e. with no spatial
features) with a $\chi^2$ distribution with one degree of freedom:
\begin{equation}
f_\chi(x) = \frac{x^{-1/2} {\rm e}^{-x/2}}{2^{1/2}\Gamma(1/2)}\ ,
\end{equation}
$\Gamma$ being the Gamma function. The latter is highly
non-Gaussian with a kurtosis [Eq.(\ref{kurtosis}) below] of 
$\gsim 10000$; no beam smoothing has been applied to it.
\begin{figure*}
\begin{center}
\epsfig{file=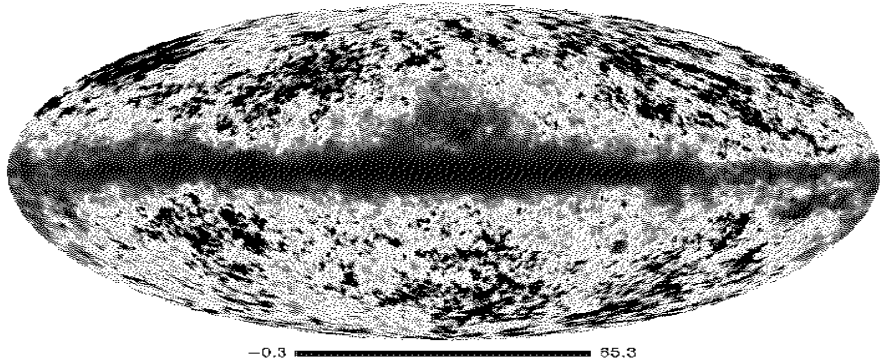,height=5.in,width=3.in,angle=90}
\caption{{\ica} reconstruction of the dust component (toy model
II). The map is plotted in a non-linear scale to emphasize the
high Galactic latitude structures, reflecting those of the CMB.}
\label{dustfull}
\end{center}
\end{figure*}
For the CMB we adopt a black body spectrum at $T_{CMB}=2.726$ K,
while the $\chi^2$ signal is assumed to have a constant antenna
temperature. We have run two cases in which the rms signal of the
latter component is, respectively, $\sim 30$ and $\sim 3$ times
larger than the CMB at 90 GHz.

As mentioned in \S~\ref{fastica} there are three kinds of
non-linear functions that approximate the neg-entropy: we have
verified that in most cases the function $g(u) = u^3$ works better
than the other two and therefore we adopt it for the following
analysis. The goodness of the reconstruction is quantified in
terms of four figures of merit: 1) the correlation coefficient
\begin{equation}
\label{rs}
r_S(x_{in},x_{out})={E[(x_{in}-\bar{x}_{in})(x_{out}-\bar{x}_{out})]
\over\sqrt{E[(x_{in}-\bar{x}_{in})^{2}]}\sqrt{E[(x_{out}-\bar{x}_{out})^{2}]}
}
\end{equation}
between the output and input data $x_{out},x_{in}$ ($E$ again denotes
the expectation value; $\bar{x}=E(x)$ represents the mean value),
2) the percentage error in the frequency scaling of the
reconstructed signal, 3) the percentage errors on higher order
moments
\begin{equation}
\label{skewness}
{\rm skewness}={E[(x-\bar{x})^{3}]\over E[(x-\bar{x})^{2}]^{3/2}}\ ,
\end{equation}
\begin{equation}
\label{kurtosis}
{\rm kurtosis}={E[(x-\bar{x})^{4}]\over E[(x-\bar{x})^{2}]^{2}}-3
\end{equation}
of the reconstructed signal, and 4) the relative residual after
re-calibration and subtraction of the corresponding input
component. These figures of merit are evaluated also for a map
with a higher number of pixels ($N_{\rm side}=128$) and a higher
angular resolution (smoothing with a $1^\circ$ FWHM beam) of the
CMB signal. 
\begin{table*}
\begin{center}
\caption{Same as in Table~1, but for a $1^\circ$ FWHM beam.}
\label{CMBchi2-1deg}
\begin{tabular}{lccccccc}
\hline
\ $N_{\rm side}$ & \multicolumn{2}{c}{$r_S$[\%]} &\multicolumn{2}{c}{$\nu$ scaling[\%]} & \multicolumn{2}{c}{moments[\%]} &
    residuals\\
                 & CMB & $\chi^2$                & CMB & $\chi^2$                       & CMB & $\chi^2$  & \\
\hline
128& $\gsim 99.99$ & $99.96$ & $\sim 0.8$ & $\sim 10^{-3}$ & $\sim 0.4$& $\sim 0.002$ & $\sim2\times 10^{-3}$\\
128$^*$&$''$& $''$ & $\sim 0.09$ & $\sim 10^{-2}$ & $''$ & $''$& $''$ \\ \hline
\end{tabular}
\end{center}
\end{table*}
As shown by Tables~\ref{CMBchi2-7deg} and \ref{CMBchi2-1deg}, the
correlation between input and output signals is extremely tight
for both components and does not vary as their intensity ratio
changes by an order of magnitude. The correlation coefficient
increases with increasing number of pixels.

The recovery of the frequency dependence of each component is also
good and obviously improves when the component becomes relatively
stronger (while the recovery of the frequency spectrum of the
other component is degraded roughly by the same factor). This
result can be understood by consideration of Eq.~(\ref{ratio}):
even when the spatial correlation is extremely good, the matrix
${\bf{W}}$ is not necessarily a robust estimator of
${\bf{A}^{-1}}$. In the case of the weaker component, better
accuracy is achieved for higher angular resolution and larger
number of pixels.
\begin{figure*}
\begin{center}
\epsfig{file=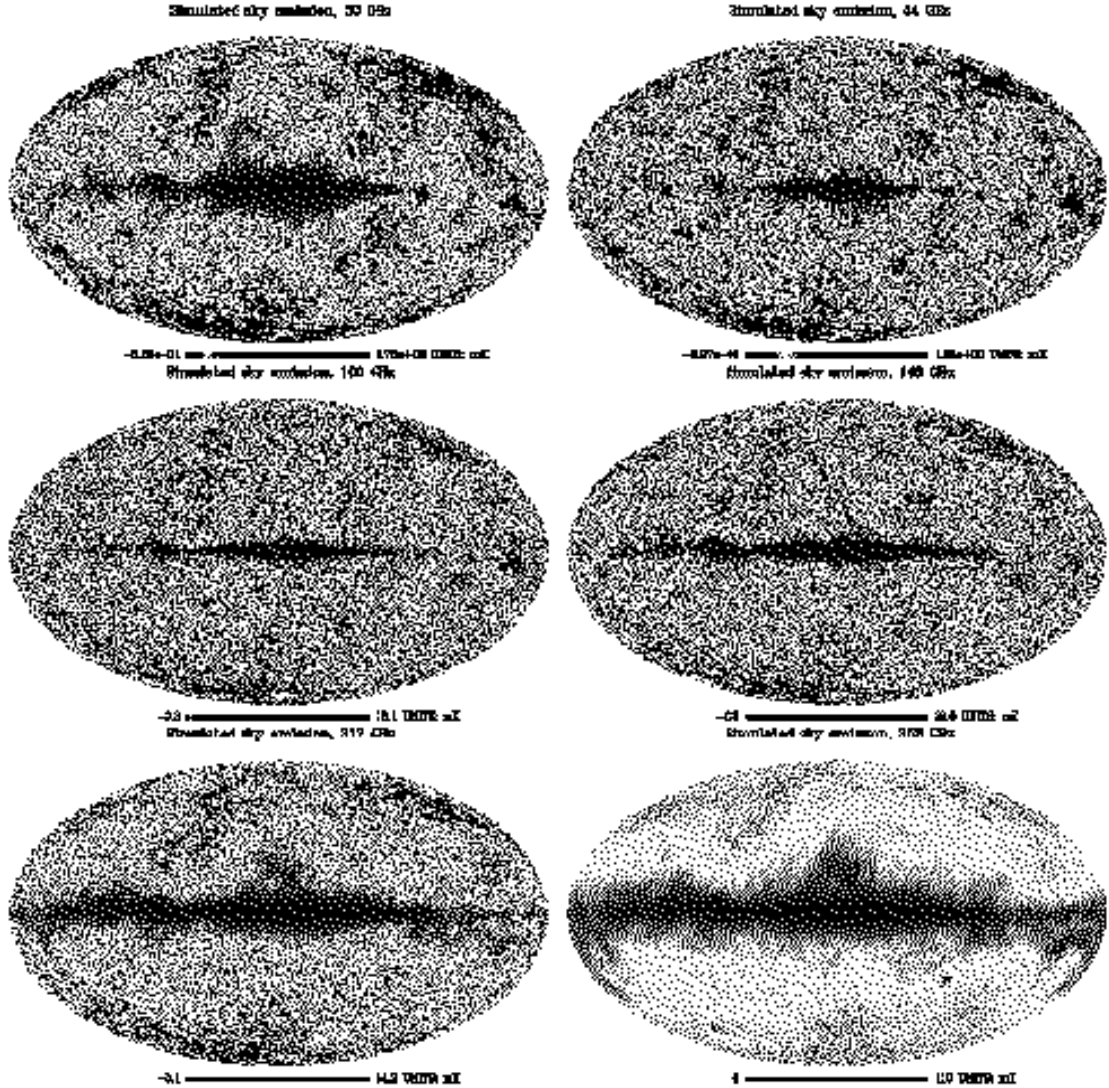,width=7.in,height=9.in}
\vskip -1.5in
\caption{Simulated skies at several {\sc Planck} frequencies.
Signals have been plotted in a non-linear scale to emphasize high
Galactic latitude structures.} \label{skies}
\end{center}
\end{figure*}

The very accurate recovery of higher order moments shows that the
algorithm, at this level of simulation, does not introduce extra
non-Gaussianity into the reconstructed signal.

Equation~(\ref{normalization}) allows us to convert the {\ica}
output into physical units. The residuals after subtraction of the
re-scaled output map from the input one, show a pattern that is
exactly the one of the other component: there is a sort of
``mirroring'' of one component into the other in the residuals.
The level of residuals depends on $N_{\rm side}$ of the map; their
$rms$ values are about $10^{-3}$ times smaller than the $rms$ of
the original map.

For fixed $N_{\rm side}$, the quality of the reconstruction is
significantly better in the case of the $1^{\circ}$ resolution,
thanks to the better sampling of the CMB statistics.


\subsection{Toy Model II}
\label{toy2}
\begin{table*}
\begin{center}
\caption{Figures of merit for the reconstruction of CMB and dust
emission convolved with a Gaussian beam with $1^\circ$ FWHM, for
different Galactic latitude cuts. The $^\dag$ indicates a case in
which the dust signal has been randomly distributed all over the
sky. Residuals are evaluated only for the CMB map.}
\label{CMBdust1deg}
\begin{tabular}{lcccccccc}
\hline
\ $N_{\rm side}$ & $b_{\rm cut}$  &\multicolumn{2}{c}{$r_S$[\%]} &\multicolumn{2}{c}{$\nu$ scaling[\%]} & \multicolumn{2}{c}{moments[\%]} &
    residuals\\
               &  & CMB & dust                & CMB & dust                       & CMB & dust  & \\
\hline
128& all sky           & $\sim  99.97$&$76.46$&$\sim 22$  & $\sim 0.3$&$\sim 6$  &$\sim 0.3$ &$0.02$\\
128& $|b| > 25^\circ$ & $\gsim 99.99$&$99.88$&$\sim 0.3$ & $\sim 4.5$&$\sim 0.7$&$\sim 0.15$&$0.008$ \\
128& $|b| \le 25^\circ$ & $\sim  99.96$&$97.67$&$\sim 11$  & $\sim 0.2$&$\sim 0.3$&$\sim 0.03$&$0.02$ \\
128$^\dag$ & all sky   & $\gsim 99.99$&$99.99$&$\sim 0.4$ & $\sim 0.001$& $\sim 0.2$ & $\sim 0.06$ & $\sim 10^{-4}$ \\
\hline
\end{tabular}
\end{center}
\end{table*}
We now replace the un-physical $\chi^2$ signal with dust emission,
modelled scaling to 53 and 90 GHz, with the spectral shape of
Eq.~(\ref{dustscaling}), the maps by Schlegel et al.~(1998). The
CMB plus dust emission maps are then convolved with a Gaussian
beam of $1^\circ$ FWHM.  We adopt $N_{\rm side}=128$.  The main
difference with the previous model is that now we have inserted a
signal (dust emission) with a strong spatial structure. We have
considered this foreground, at frequencies where Galactic
synchrotron and free-free emissions should dominate, because its
spatial distribution is known at high resolution (6$'$) and we
wanted to explore the effect of its markedly non-uniform sky
distribution, highly concentrated on the Galactic plane.

The power-law function, $g(u)=u^{3}$, still works better than the
others. A summary of the results, in terms of the usual figures of
merit, is in Table~\ref{CMBdust1deg}. We note that, in the case of
the all-sky analysis (first row of the Table), while the
reconstructed CMB map is very well correlated with the input one,
this is not the case for the dust. A significant degradation of
the reconstructed dust map is clearly visible in
Fig.~\ref{dustfull}, representing the output dust map, showing
some CMB spatial features at high Galactic latitudes. Furthermore
the derived frequency dependencies of both signal, but
particularly that of the CMB, deviate significantly from the
expected one. In addition, when the CMB map is converted into
physical units, a residual map showing dust spatial structures is
present with an $rms$ only 50 times smaller than the CMB $rms$,
which is much worse than the result we obtained in Section 4.1.

The main reason of these results is that the dust is highly not 
uniformly distributed across the sky. Indeed, 
as noted above, the recovery of the frequency dependence, i.e. of 
the separation matrix, is poor if one component is much weaker than 
the other; the situation worsens if the ratio of the two components 
is strongly varying across the map, as is the case here: dust dominates 
on the Galactic plane, while the CMB dominates at high Galactic 
latitudes. In addition, dust has different statistics on and off 
the Galactic plane, as Table.~\ref{distri} shows. 
In order to check this point, we made an experiment 
separating CMB and an artificial ``sky-uniform" dust map generated 
by randomly distributing its signal over all the sky 
(simply randomly shufflying the pixels so that the Galactic 
plane structure is destroyied): the quality of the reconstructed 
maps becomes extremely good (fourth row in Table~\ref{CMBdust1deg}). 

These simulations show that a better reconstruction can be
achieved for uniformly distributed sky signals. In the present 
problem, this can be obtained by analyzing separately regions around 
the Galactic plane and at high Galactic latitudes (second and third 
row of Table~\ref{CMBdust1deg}), where the Galactic emission has very
different spatial distributions (see Table.~\ref{distri}).

\begin{table}
\begin{center}
\caption{Skewness and kurtosis of dust emission and dust/CMB $rms$
ratio for sky regions at low and high Galactic latitudes.}
\label{distri}
\begin{tabular}{lcccc}
\hline
\ $b_{\rm cut}$ & skewness & kurtosis & \multicolumn{2}{c}{dust/CMB} \\
 & & & 53~GHz & 90~GHz \\
\hline
\ $|b| > 25^\circ$   & 0.01 &  -2.97 & 0.095 & 0.23\\
\ $|b| \le 25^\circ$ & 7.44 &  139.3 & 13.19 & 5.49\\
\hline
\end{tabular}
\end{center}
\end{table}
\subsection{Effect of Instrumental Noise}
\label{toy3} We now add instrumental noise to the toy model
considered in the previous sub-section. The noise is assumed to be
uniform across the sky and such that the average signal-to-noise
ratio at high Galactic latitudes is $\simeq 2$. Again, the all-sky
reconstruction is unsatisfactory. However, if we analyze
separately regions within $|b|=25^\circ$ and at higher Galactic
latitudes, we recover the CMB frequency scaling with a 0.5\% error
at high Galactic latitudes and that of dust emission with the same
error at low latitudes. The signal-to-noise ratio at high Galactic
latitudes is $\simeq 2$ for the CMB (as in the input map) and
$\simeq 1$ for dust. At low Galactic latitudes the signal-to-noise
ratio is still $\simeq 2$ for the CMB, and $\simeq 12$ for dust
emission.

\section{{\ica} and {\sc Planck} simulated maps}
\label{planck} We simulate all-sky maps adopting the HEALPix
pixelization scheme \cite{GORSKI}. The maps are in antenna
temperature and, as described in detail below, include thermal
dust, free-free, and synchrotron emissions, on top of CMB
fluctuations.
\begin{figure*}
\begin{center}
\epsfig{file=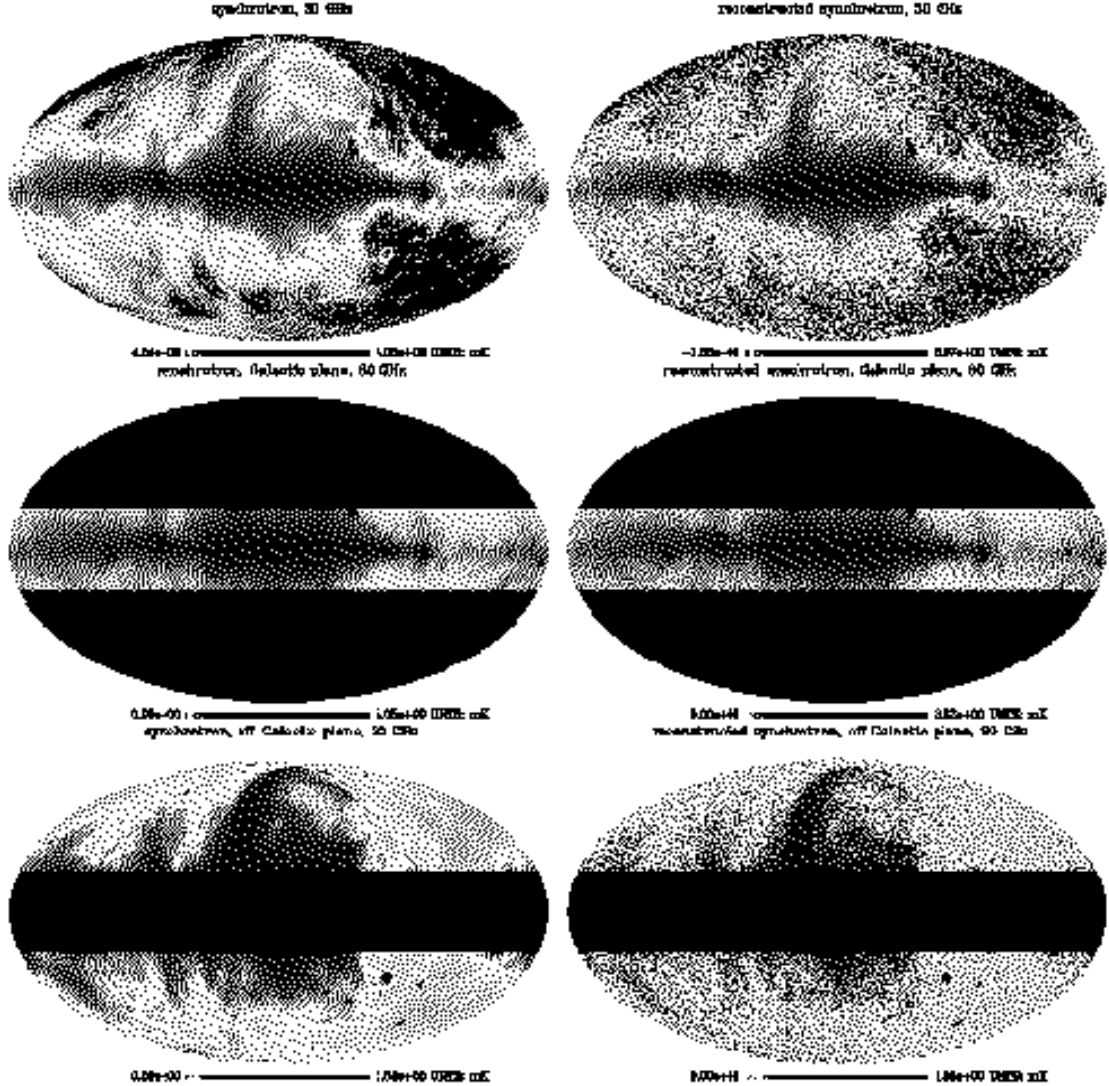,width=7.in,height=9.in}
\vskip -1.5in
\caption{Input (left) and output (right) synchrotron maps for case
1,  at 30 GHz. Maps have been
plotted in a non-linear scale to emphasize high Galactic latitude
features.} \label{syn1}
\end{center}
\end{figure*}
For the CMB we assume a black body spectrum with
$T_{CMB}=2.726\,$K, so that fluctuations in antenna temperature,
$\delta T_{A, CMB}$ are related to brightness temperature
fluctuations $\delta T_{CMB}$ by:
\begin{equation}
\delta T_{A, {\rm CMB}}(\nu)= {x^{2}e^{x}\over (e^{x}-1)^{2}}
\delta T_{CMB} \ , \label{CMBscaling}
\end{equation}
where $x=h\nu/k T_{CMB}$. We have adopted brightness temperature
fluctuations, $\delta T_{CMB}$, corresponding to a flat
$\Lambda$CDM cosmology with $\Omega_{CDM}=0.35$, $ \Omega_b=0.05$,
$\Omega_{\Lambda}=0.6$ and $h=0.65$, three massless neutrino
families and an initial scale-invariant Gaussian spectrum.
Galactic foreground emissions are described in detail below.

The CMB and foreground maps are then summed up together with
simulated white noise maps, generated by assuming the rms noise
fluctuation level expected for the {\sc Planck} channels (see
Table \ref{tablenoise}). The simulated skies can be seen in Fig.
\ref{skies}.
\begin{table*}
\begin{center}
\caption{Assumed parameters for the Planck channels considered
here. Sensitivities, for a 14 months mission, are in antenna
temperature.}
\begin{tabular}{l c c c c c c c c}
\hline
& \multicolumn{3}{c}{LFI} & \multicolumn{4}{c}{HFI} \\
\hline
frequency (GHz) &
30 & 44 & 100 & 100 & 143 & 217 & 353 \\
Angular resolution ($'$) &
33.6 & 22.9 & 10  & 9.2 & 7.1 & 5.0 & 5.0 \\
$\Delta$T ($\mu$K) &
5.1 & 7.8 & 12.4 & 4.7& 4.0 & 3.5 & 3.2 \\
\hline
\end{tabular}
\label{tablenoise}
\end{center}
\end{table*}

Given our sky composition, three cases exploiting three subsets of
{\sc Planck} frequency channels are considered. The first case
(case 1) makes use of the two lowest frequency channels (30 and 44
GHz) assuming a mixture of CMB plus the dominant diffuse Galactic
foreground, which is assumed to be the synchrotron emission. Note
that a significant contribution from Galactic free-free should be
present in the real sky, at least at low Galactic latitudes, but a
good template for it is still missing. For the moment, we neglect
it. We apply {\ica} to all-sky maps as well as to regions of high
($|b|>20 ^{\circ}$) and low ($|b|<20^{\circ}$) Galactic latitude.

The second application (case 2) is more CMB oriented: it refers to the
{\sc Planck}  ``cosmological" channels at 100, 143 and 217 GHz, where
foregrounds are at a level comparable with noise; in this case, we apply
{\ica} only to all sky maps.

The third case (case 3) considers again two signals and two
channels, but we move to higher frequencies, 217 and 353 GHz,
where the CMB and our thermal Galactic component, almost
coincident with dust, dominate over synchrotron, which was
neglected. Again we apply {\ica} to all sky maps, to the region
$|b|<20^{\circ}$ and to the complementary region at high Galactic
latitudes.

Before to go to the results, let us describe briefly the
methodology used. Simulated maps were originally produced with
pixels of size 3.5$'$, corresponding to $N_{side}=1024$ in the
HEALPix pixelization scheme, and such high-resolution maps
should be taken as ``reference'' skies. However, we can note from
Table \ref{tablenoise} that the highest frequency channels have
5$'$ resolution, which in principle requires a smaller pixel size
to Nyquist sample the beam-width. Since in this preliminary study
we are neglecting the astrophysical signals relevant on the
smallest scales (such as extragalactic point sources), we restrict
our analysis to multipoles $\ell \lsim 2000$, for which pixels of
3.5$'$ are enough.

The reference maps have been smoothed with the appropriate FWHM of
{\sc Planck} channels, and the corresponding instrumental noise
was added. Due to {\it Assumption 3} in Sect. \ref{formalization},
the {\ica} algorithm requires input maps having the same angular
resolution. This situation has been achieved by reducing, for each
case considered, the effective working angular resolution
FWHM$_{\rm work}$ to the worst one. This implies an
extra-smoothing of the observed (signal $+$ noise) maps having
FWHM$<$FWHM$_{\rm work}$. Note that the {\sc Planck} nominal noise
(where the noise is assumed uniformly distributed on the sky)  is
generally given per resolution element corresponding to beams of
different size, depending on the channel. This extra-smoothing
process required by {\ica} changes the noise rms with respect to
the value reported in Table \ref{tablenoise}: the new rms values
are used to construct the ${\bmath{\Sigma}}$ matrix in
Eq.~(\ref{qwhit1}). Simulations on cases 2 and 3 have been
performed maintaining $N_{side}=1024$, while in the case 1 we
worked with $N_{side}=256$ because of the smaller angular
resolution of the lowest {\sc Planck} channels.

\begin{table*}
\begin{center}
\caption{Percentage errors on {\ica} frequency scaling
reconstruction.}
\begin{tabular}{l c c c c}
\hline \hline {\bf Considered frequencies (GHz):} & {\bf 30/44} &
{\bf 100/143}  &
{\bf 143/217} & {\bf 217/353} \\
\hline \hline
{\bf Case 1, all-sky} & & & &\\
\hline
CMB & 1 & & &\\
Synchrotron & 3 & & &\\
\hline
 {\bf Case 1, $|b|>20 ^{\circ}$}  & & & &\\
\hline
CMB & 1 & & &\\
Synchrotron & 0.3 & & & \\
\hline
{\bf Case 1, $|b|<20 ^{\circ}$}  & & & &\\
\hline
CMB & 2 & & &\\
Synchrotron & 3 & & &\\
\hline
 {\bf Case 2, all-sky } & & & &\\
\hline
CMB & & 1 & 0.6 &     \\
Thermal component & & 0.4 & 0.2  &   \\
Synchrotron & & 75 & 84  &   \\
\hline
 {\bf Case 3, all-sky }  & & & &\\
\hline
CMB & & & & 46  \\
Thermal component & & & & 0.3  \\
\hline
 {\bf Case 3, $|b|> 20 ^{\circ}$} & & & & \\
\hline
CMB & & & & 12\\
Thermal component & & & & 0.7\\
\hline
{\bf Case 3, $|b|<20 ^{\circ}$} & & & & \\
\hline
CMB & & & & 64\\
Thermal component & & & & 0.03\\
\hline
\end{tabular}
\label{tablescalings}
\end{center}
\end{table*}

We quantify the quality of {\ica} outputs with reference to the
percentage errors on frequency scaling recovery, on normalization,
and on the angular power spectrum. For each application we
estimated the relative error on the scaling reconstruction between
nearby frequencies, by comparing the ratios of the ${\bf{W}}^{-1}$
matrix elements in Eq.~(\ref{scalingsjk}) with their input values.
The results, for all the cases, are shown in Table
\ref{tablescalings}.
\begin{figure*}
\begin{center}
\epsfig{file=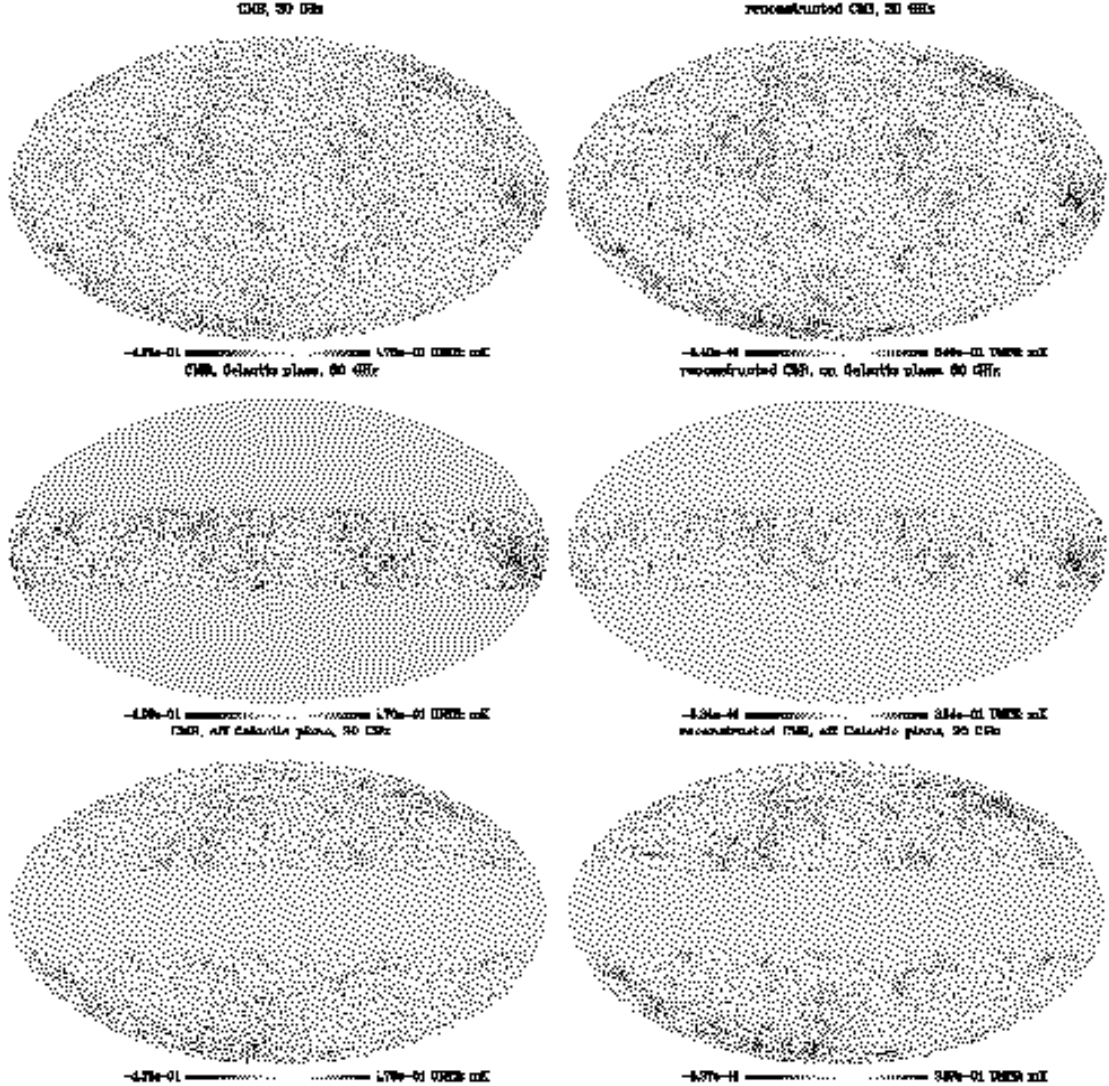,width=7.in,height=9.in}
\vskip -1.5in
\caption{Input (left) and output (right) CMB maps for case 1,
 at 30 GHz.} \label{CMB1}
\end{center}
\end{figure*}
The normalization allows us to measure in some detail how the
noise affects the separation matrix, since as we have seen in
\S~2, Eq.~(\ref{normalization}), the normalization of the {\ica}
outputs to physical units involves a direct product of an element
of the inverted separation matrix by a {\ica} output. The angular
power spectrum of each physical component is estimated using the
{\ica} output maps as follows. At each frequency, and for each
signal, the observed $C_{\ell}$ are the sum of signal and noise
contributions:
$$
C_{\ell}^{map}= C_{\ell} W(\ell)+C_{\ell}^{noise}
$$
where 
the filter function
\begin{equation}
\label{Wl} W(\ell)={\rm exp}{[-\ell(\ell+1)({\rm
FWHM/rad})^{2}/(2\log{2})]}
\end{equation}
describes the effect of the beam shape on multipoles and
$C_{\ell}^{noise}$ represents the noise contamination, which is
additive only if noise and signal are perfectly uncorrelated. Our
guesses about the noise rms allow us to Monte Carlo simulate the
$C_{\ell}^{noise}$. The power spectra of {\ica} output noisy maps,
obtained with the code {\tt Anafast} (HEALPix package), should
therefore be a good estimate of the quantities $[C_{\ell}
W(\ell)+C_{\ell}^{noise}]$ for the signals observed. We can invert
the previous relation and estimate the power spectra of the
signals, since we know $W(\ell )$ and have a guess for
$C_{\ell}^{noise}$:
\begin{equation}
C_{\ell} =
\left[ {C_{\ell}^{map}-C_{\ell}^{noise} \over W(\ell)} \right] \ \ .
\label{clrecovery}
\end{equation}
This is the relation used in the following analysis, to evaluate the
performances of {\ica} separation in terms of the angular power spectrum.

\subsection{Case 1}
Components to be recovered are CMB and synchrotron which are
assumed to dominate the sky emission at 30 and 44 GHz. The
synchrotron emission template is the 408 MHz map of Haslam et al.
(1982), extrapolated to the considered frequencies assuming a
uniform spectral index $ \beta_{\rm syn}=-2.9$. On account of the
poor resolution of the Haslam map ($\sim 0.85^{\circ}$) we
artificially add Gaussian small scales fluctuations with a power
spectrum $P_{\ell} \propto \ell^{-3}$.

\begin{figure*}
\begin{center}
\epsfig{file=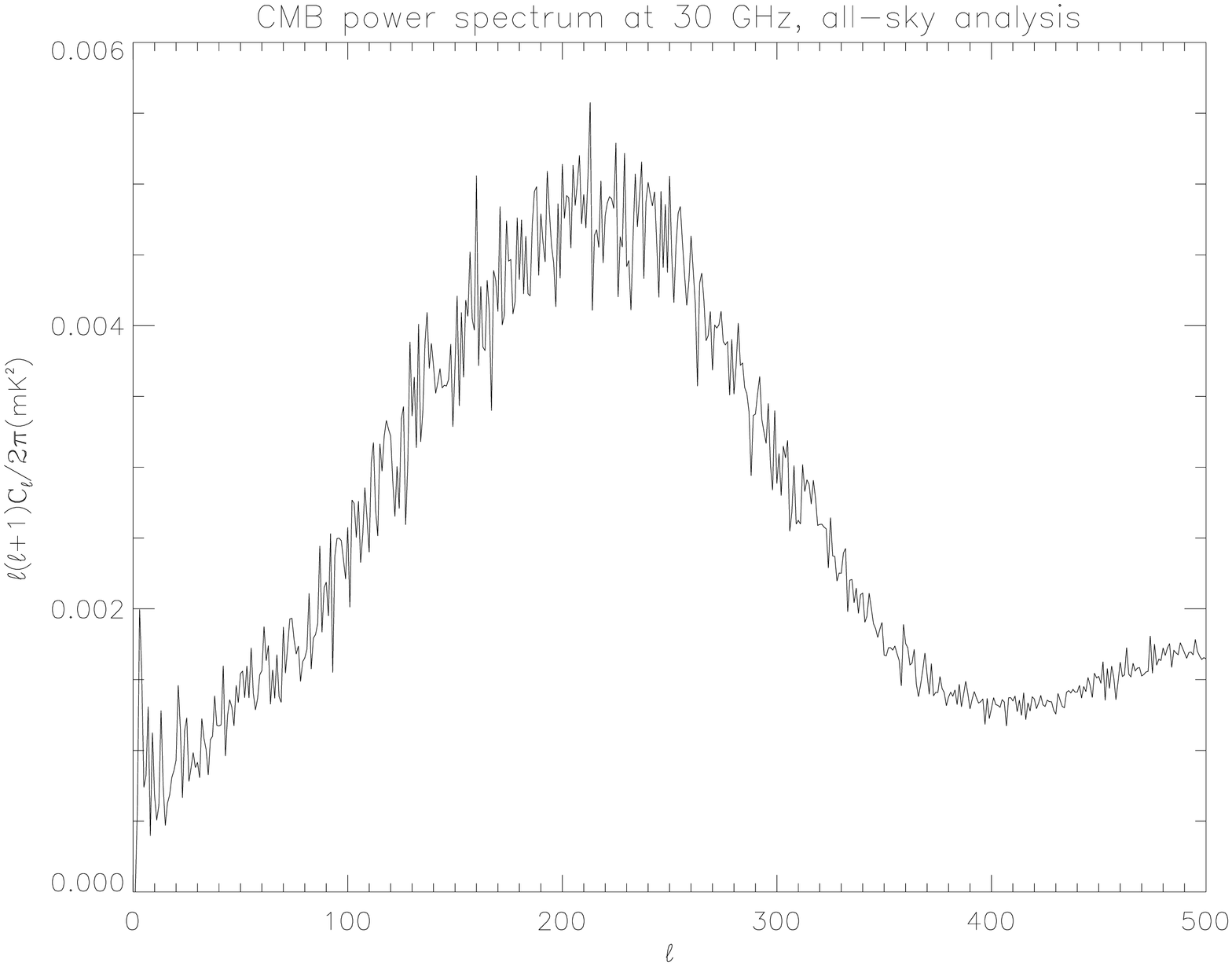,height=3.in,width=2.in,angle=90}
\hskip .2in
\epsfig{file=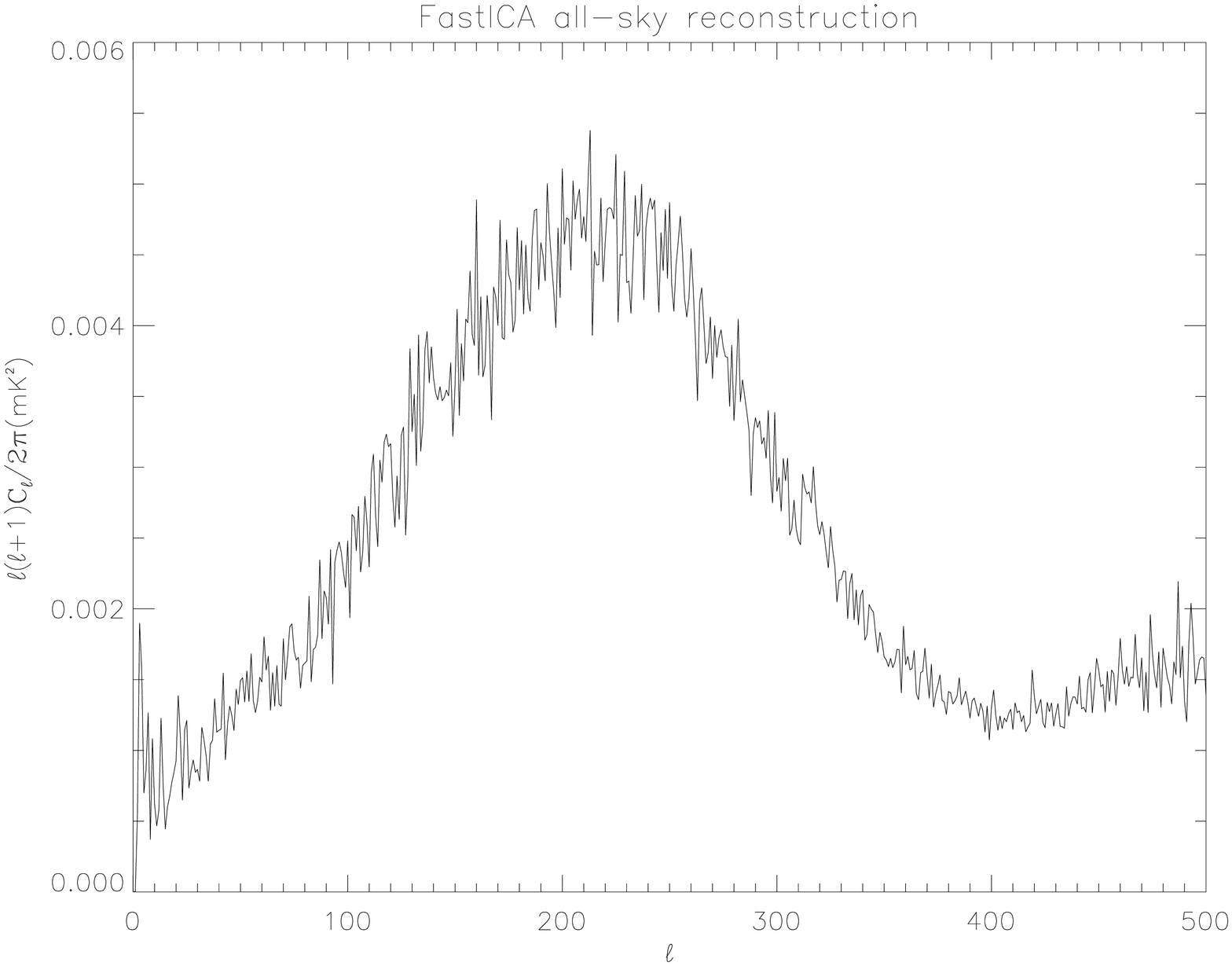,height=3.in,width=2.in,angle=90}
\vskip .3in
\epsfig{file=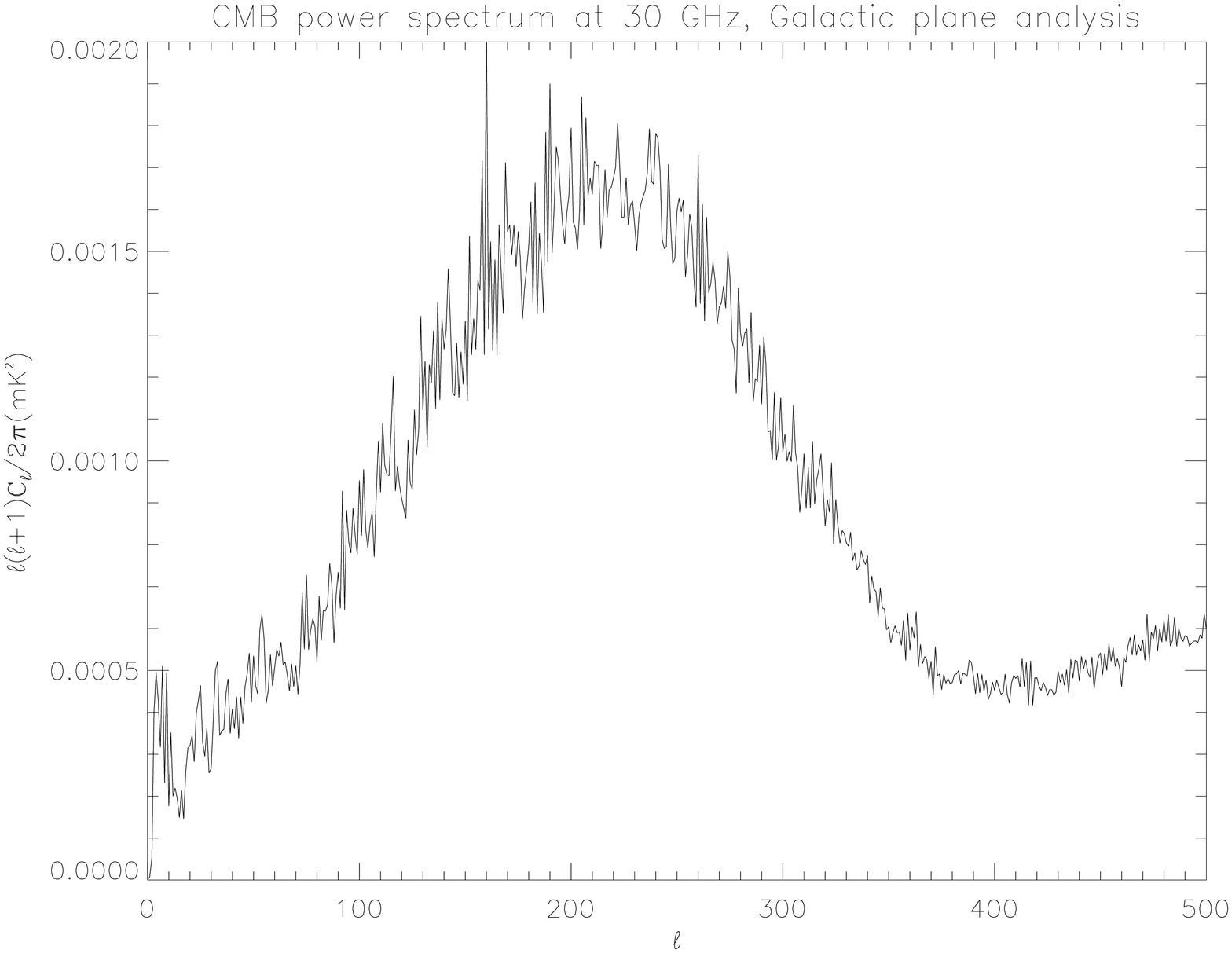,height=3.in,width=2.in,angle=90}
\hskip .2in
\epsfig{file=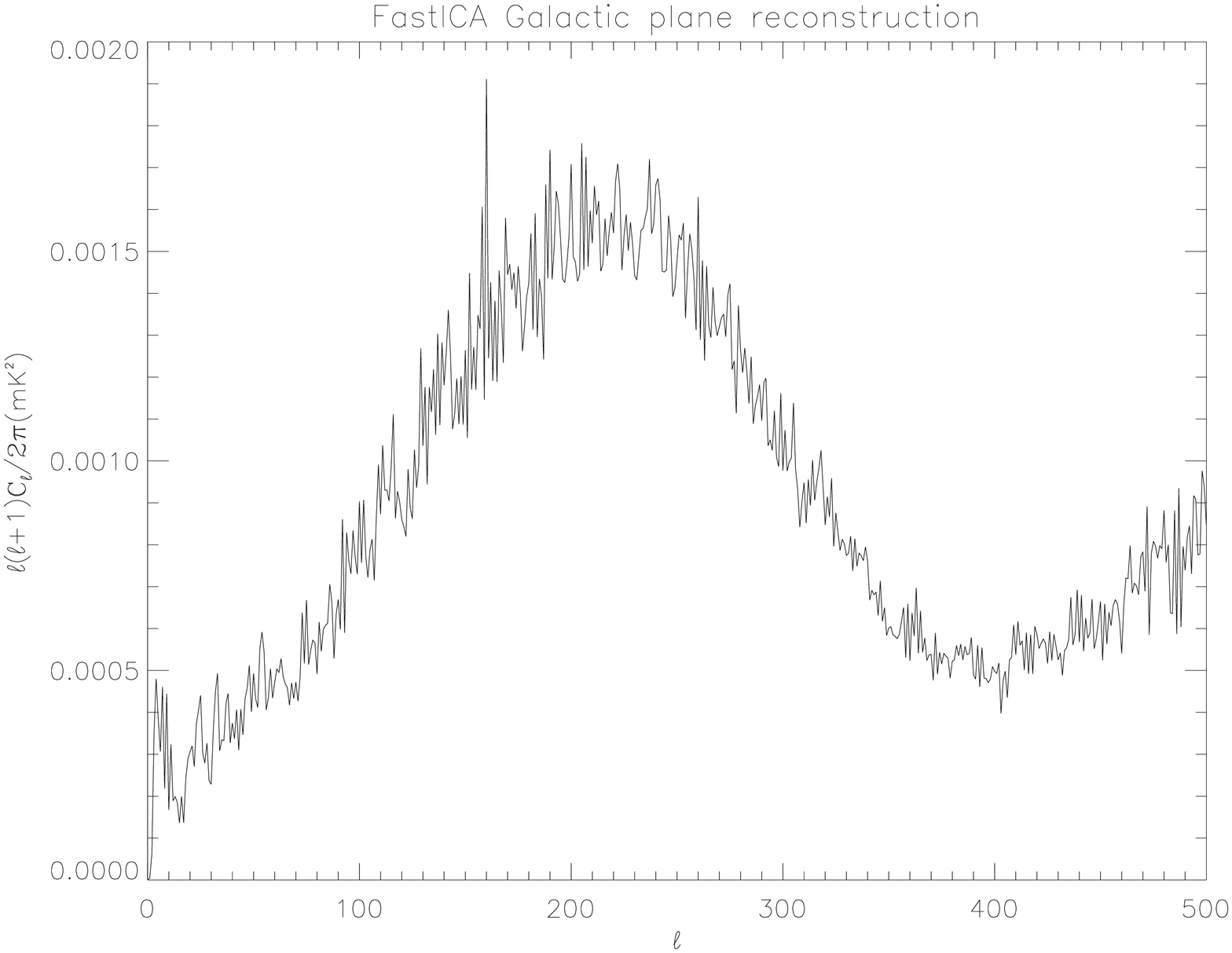,height=3.in,width=2.in,angle=90}
\vskip .3in
\epsfig{file=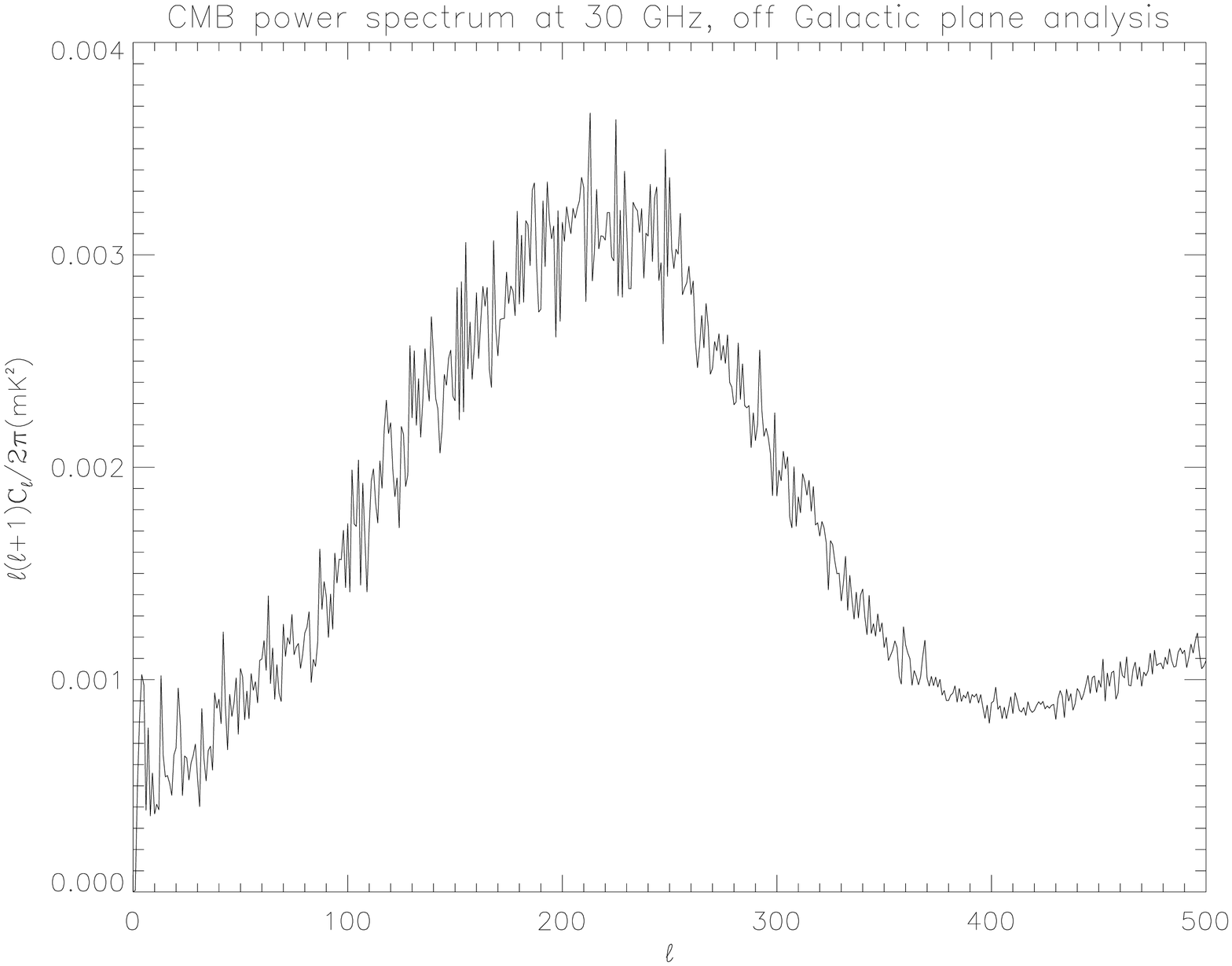,height=3.in,width=2.in,angle=90}
\hskip .2in
\epsfig{file=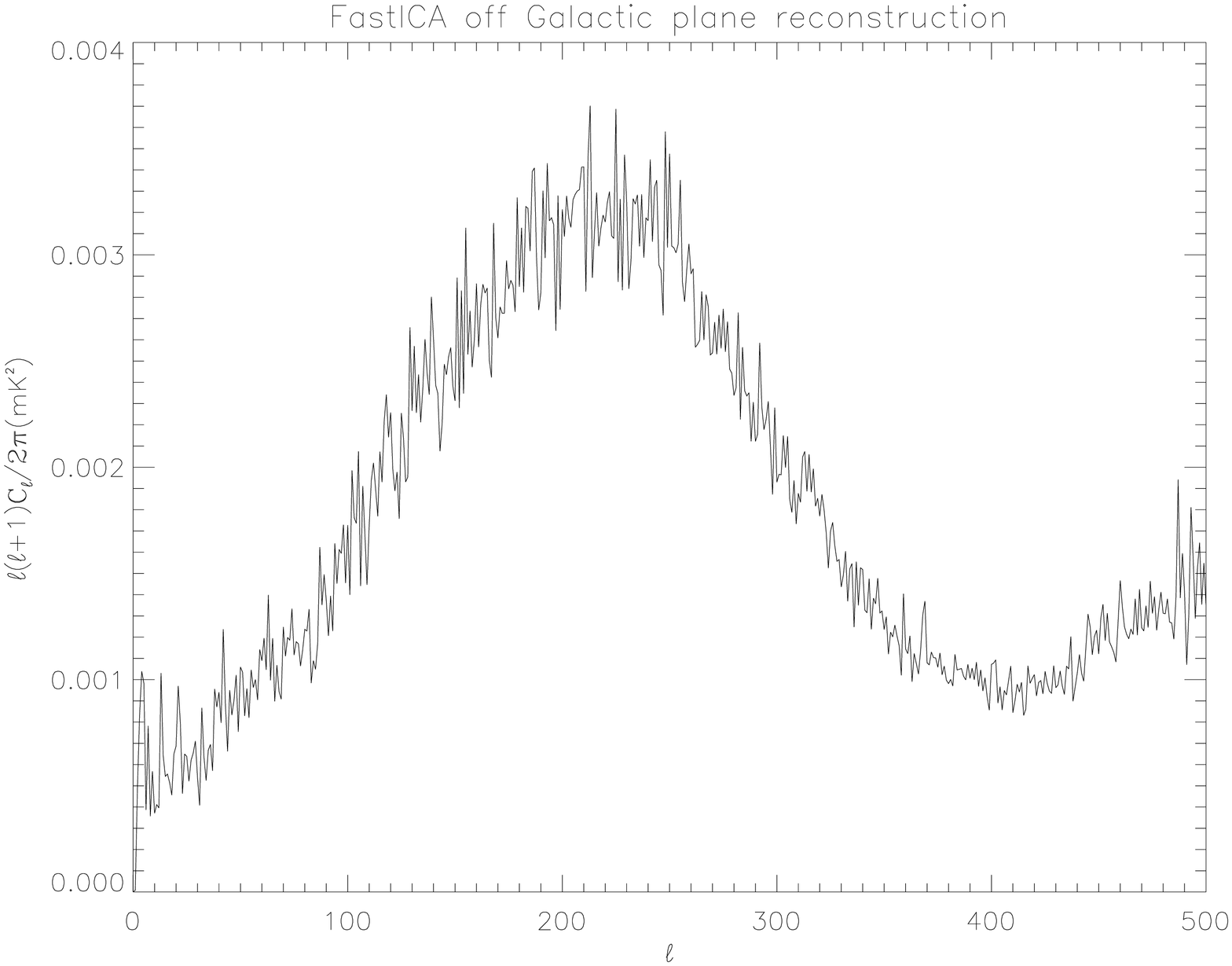,height=3.in,width=2.in,angle=90}
\vskip .3in \caption{Input (left) and output (right) CMB angular
power spectra for case 1, at 30 GHz. From top to bottom: all-sky,
low, and high Galactic latitude {\ica} results, respectively.}
\label{cl_CMB1}
\end{center}
\end{figure*}

As already mentioned, we have applied the {\ica} algorithm to the
whole sky, as well as to regions around the Galactic plane
($|b|<20^{\circ}$) and of high Galactic latitude
($|b|>20^{\circ}$). The 44 GHz noisy map was smoothed to a ${\rm
FWHM}_{\rm work}=33.6'$, corresponding to the {\sc Planck}
resolution at 30 GHz.
\begin{figure*}
\begin{center}
\epsfig{file=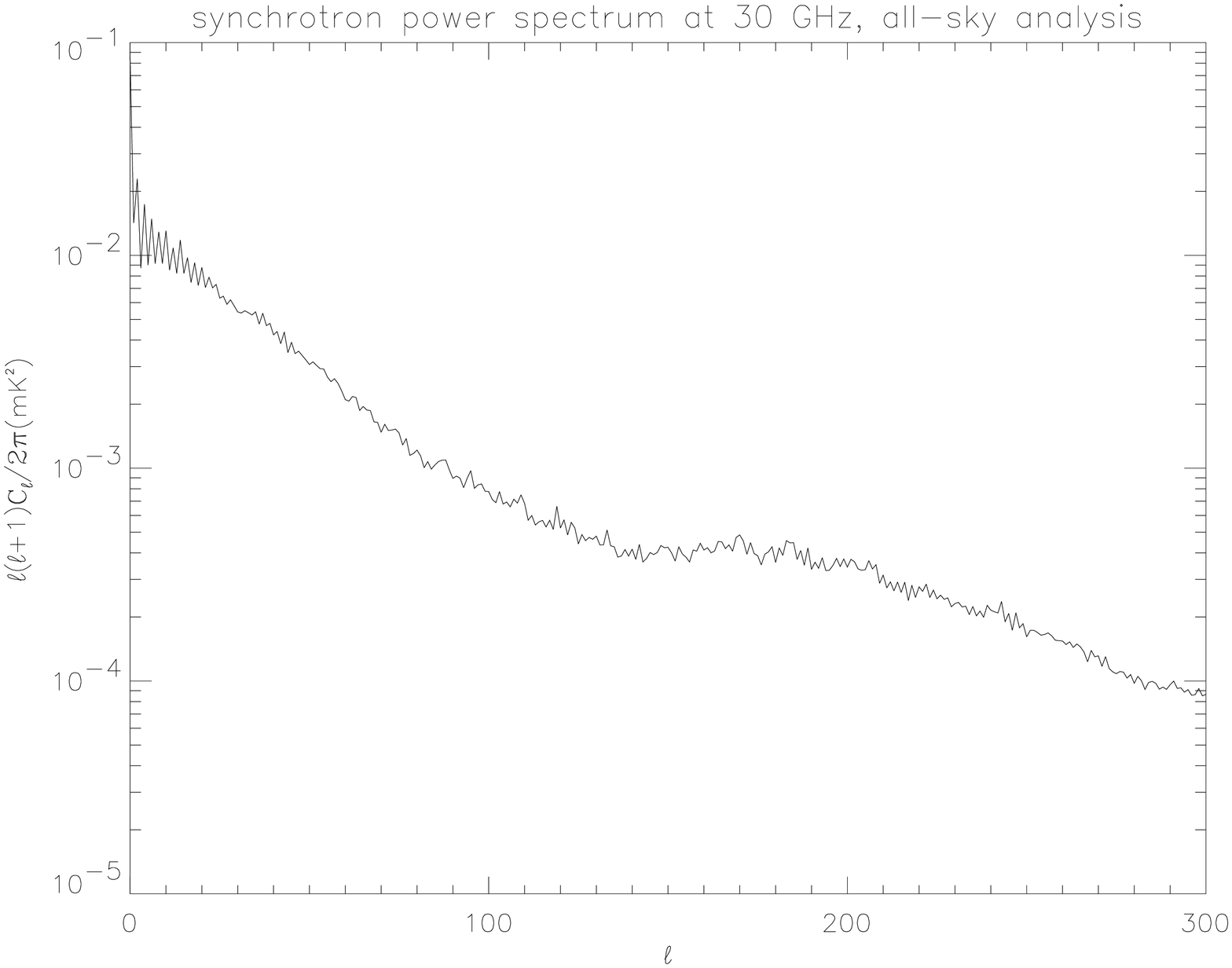,height=3.in,width=2.in,angle=90}
\hskip .2in
\epsfig{file=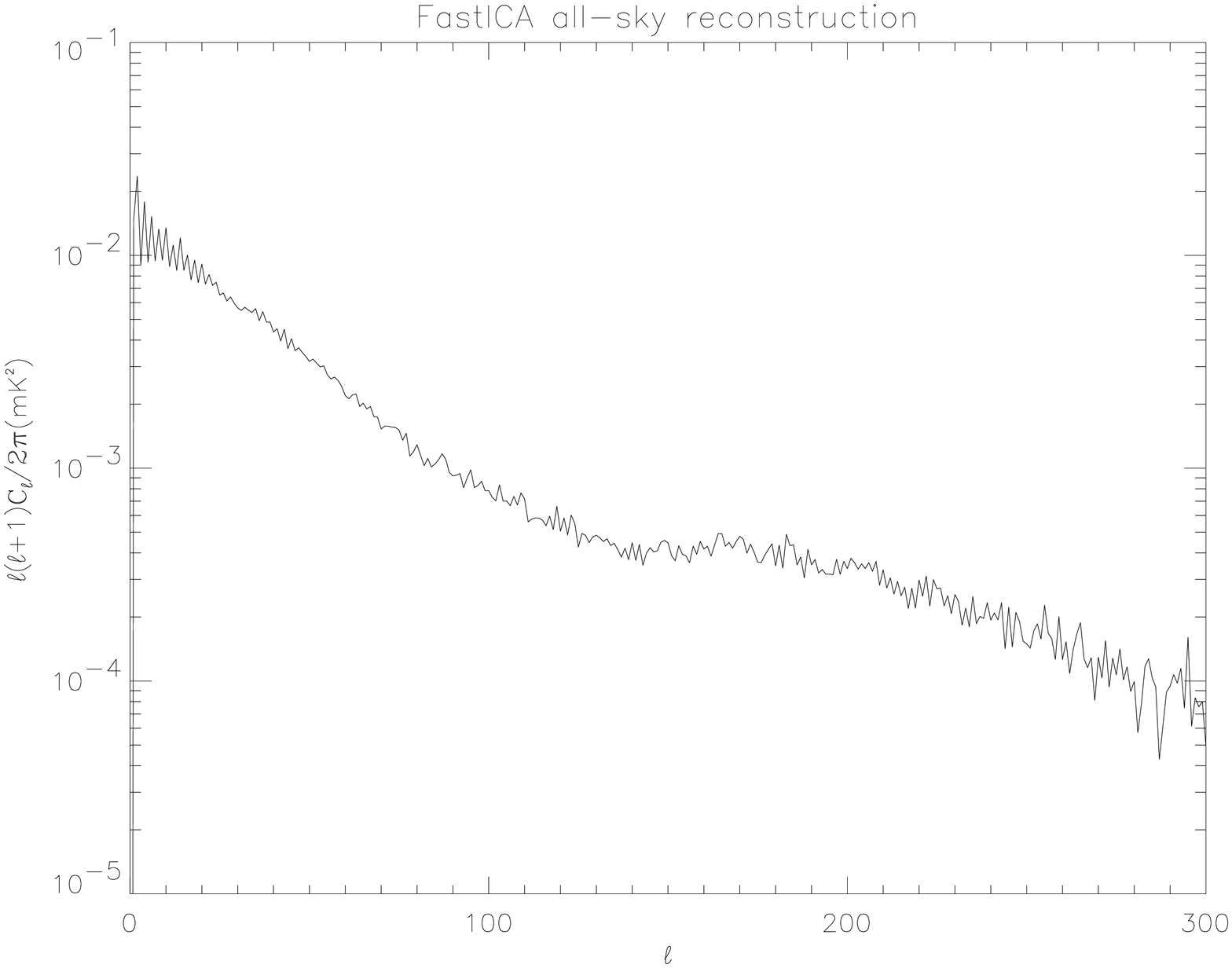,height=3.in,width=2.in,angle=90}
\vskip .3in
\epsfig{file=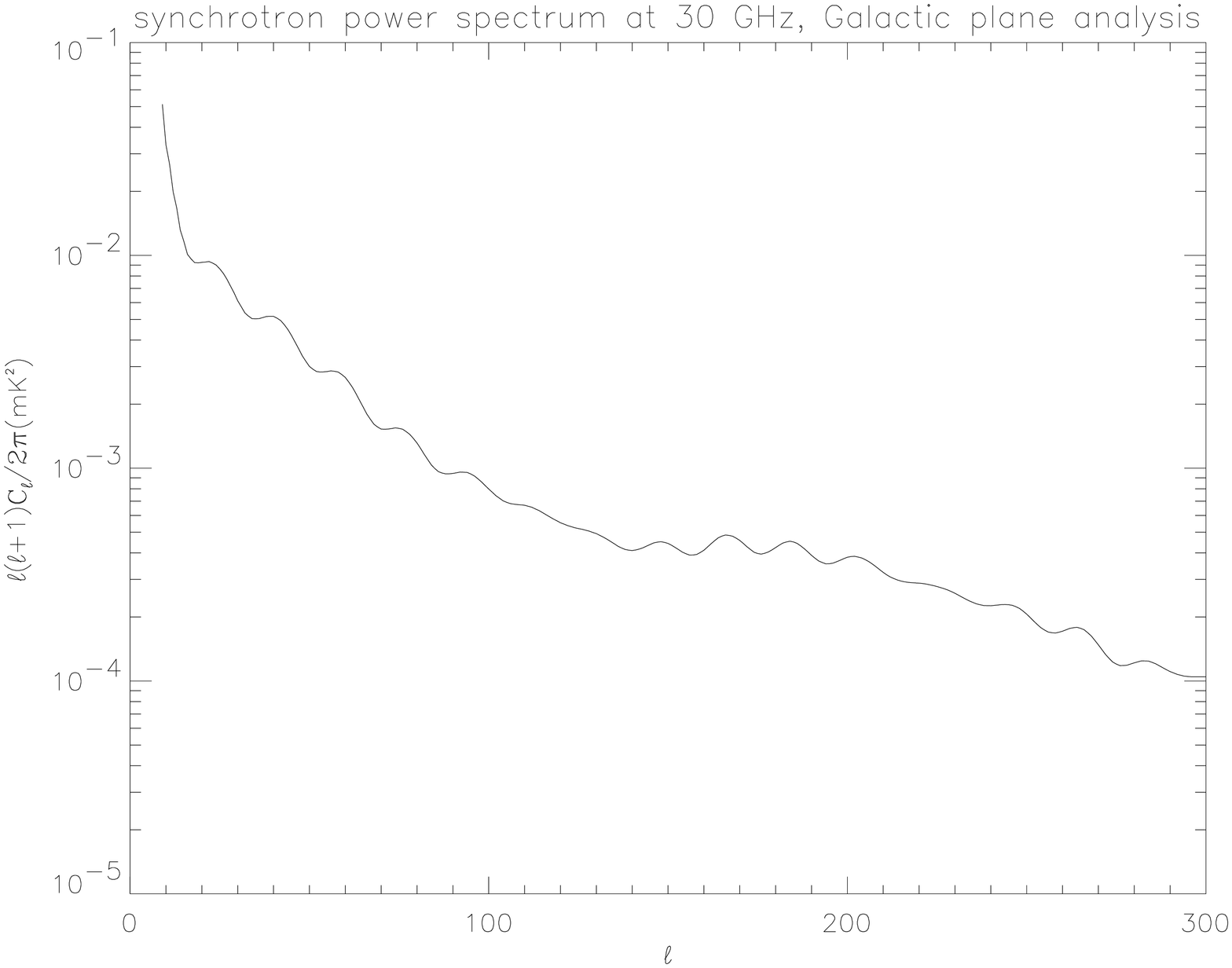,height=3.in,width=2.in,angle=90}
\hskip .2in
\epsfig{file=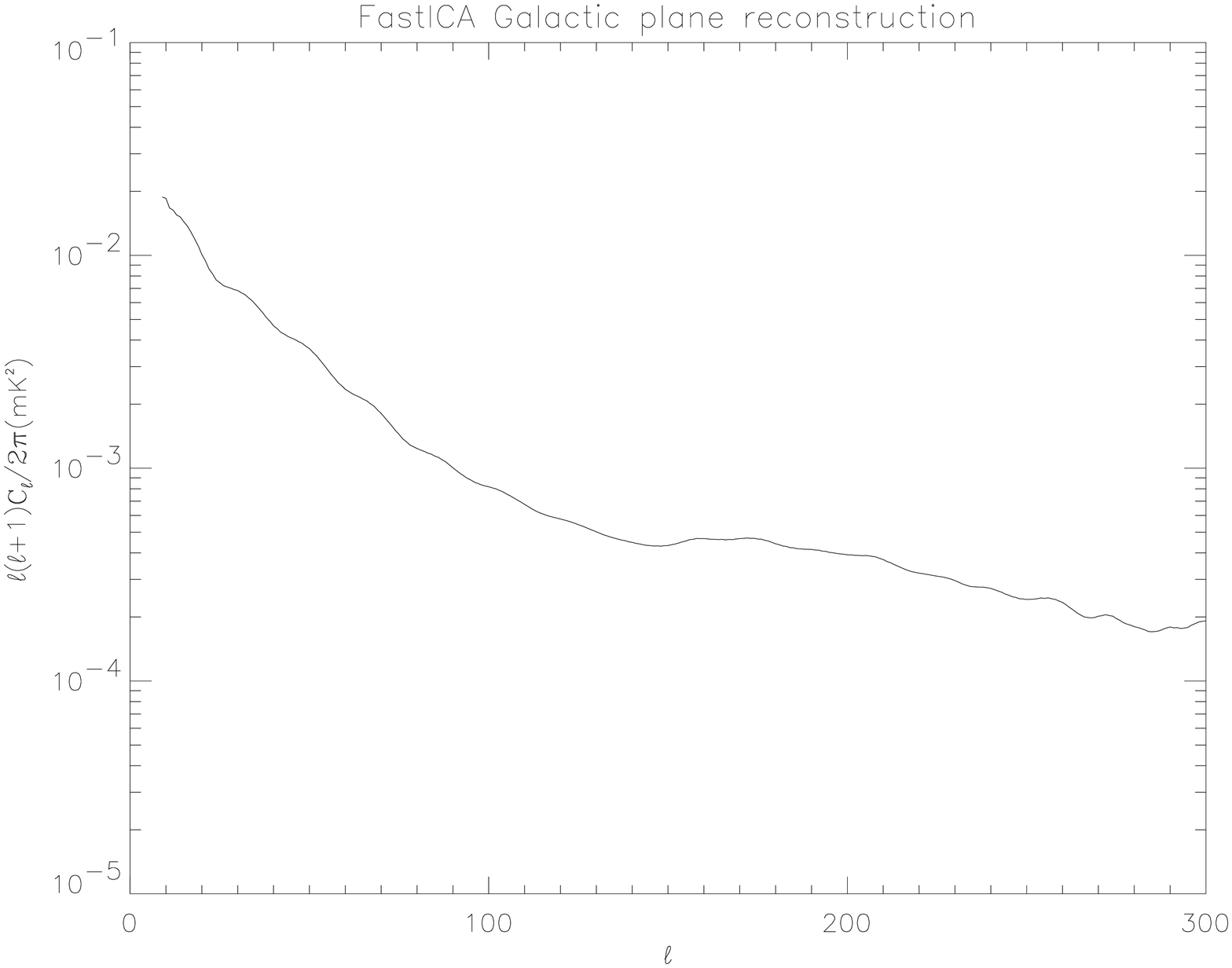,height=3.in,width=2.in,angle=90}
\vskip .3in
\epsfig{file=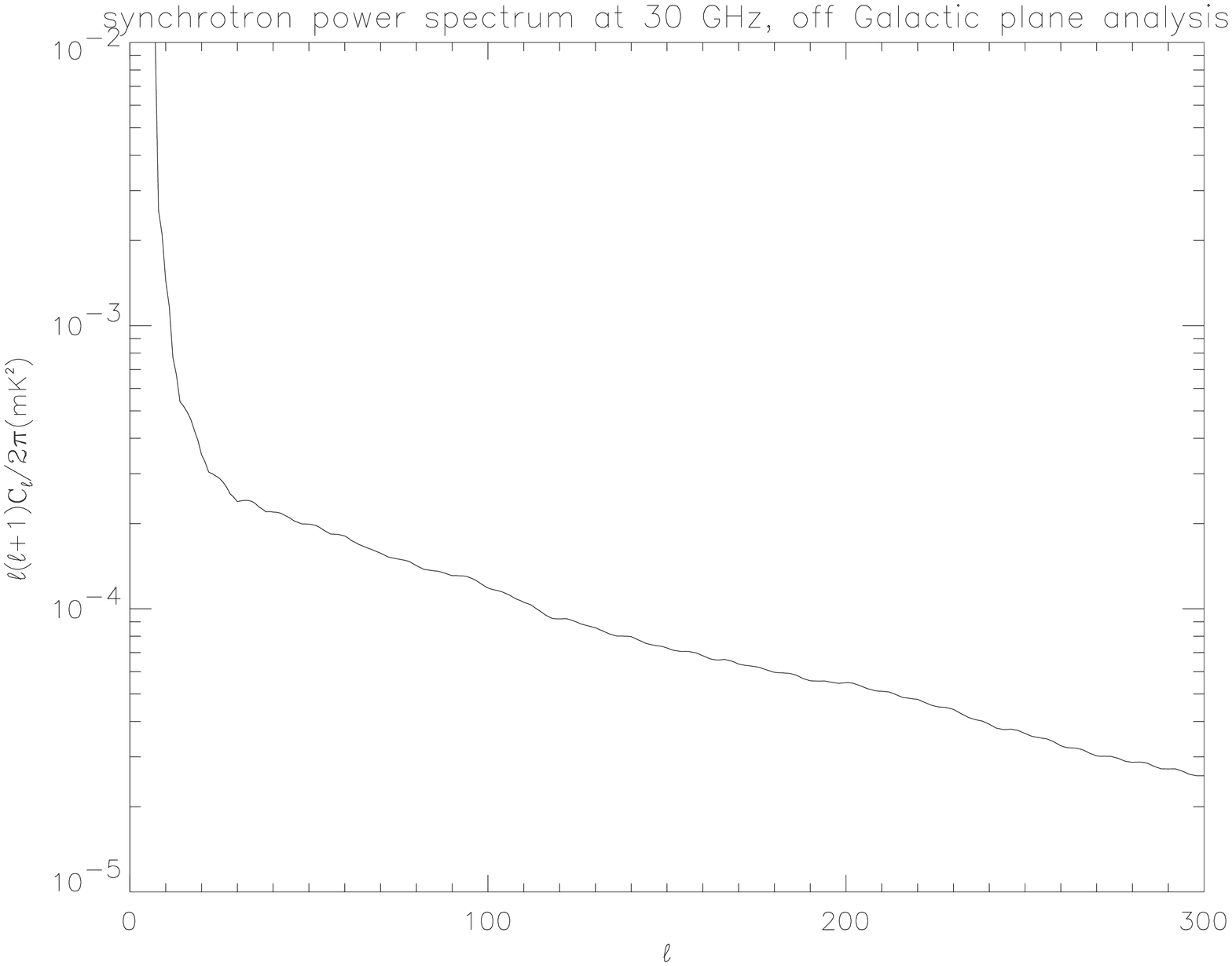,height=3.in,width=2.in,angle=90}
\hskip .2in
\epsfig{file=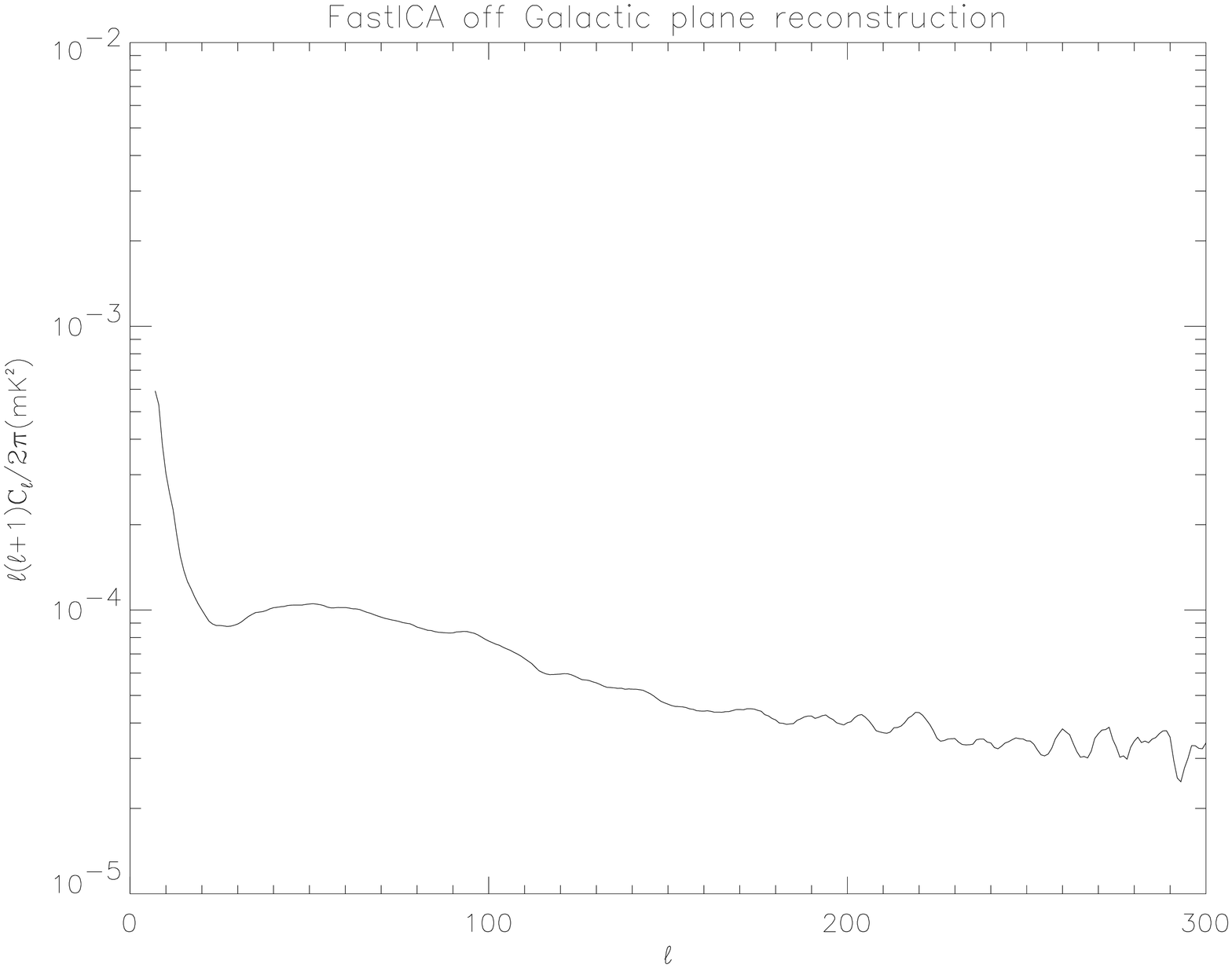,height=3.in,width=2.in,angle=90}
\vskip .3in \caption{Input (left) and output (right) synchrotron
angular power spectra for case 1, at 30 GHz. From top to bottom we
plotted all-sky, low and high Galactic latitude {\ica} results,
respectively.} \label{cl_syn1}
\end{center}
\end{figure*}
We finally re-gridded the maps with pixels of size $\sim 14'$,
(corresponding to $N_{\rm side}$=256), which is enough to Nyquist
sample the working beam size. Figures \ref{syn1},\ref{CMB1} show
input versus output maps, while figures
\ref{cl_syn1},\ref{cl_CMB1} give the corresponding power spectra,
showing, from top to bottom, the all-sky, high and low Galactic 
latitude results, respectively.

Synchrotron is reconstructed remarkably well over whole sky. The
residual noise shows up in the high $\ell$ tail of the
reconstructed spectrum. The normalization is also correctly
recovered. The reconstruction for regions at low and high Galactic
latitudes is still good, although it does not reach the quality of
the all-sky analysis. Note however that the synchrotron map has
essentially no power above $\ell\simeq 100$; this is the reason
why we plotted this signal only up to multipoles $\ell\simeq 300$
corresponding to scales slightly below $1^\circ$. When the angular
power spectrum of this signal is estimated for a fraction of the
sky, oscillations are found, which propagate down to very low
$\ell$s. Such oscillations have been smoothed by averaging
over intervals $\Delta \ell =10$. However {\sc FastICA} has
recovered the correct normalization of the reconstructed
synchrotron, as can be seen by looking at the scales in Figs.
\ref{syn1} and \ref{cl_syn1}.

The CMB is reconstructed well on all the scales down to the
adopted resolution scale, which corresponds to $\ell\simeq 500$.
Also the normalization was recovered quite correctly. 
The rise in the recovered spectra at $\ell\simeq 500$ is because 
the maps have been smoothed and then gridded before separation, as 
we explained above: this introduces extra power on the spectrum, at 
multipoles corresponding to the pixel size, which is enhanced by 
dividing by the beam window function as in Eq.(\ref{clrecovery}). 

The percentage errors
on the reconstructed frequency scalings, reported in
Table~\ref{tablescalings}, are at the percent level for both
components.

\subsection{Case 2}
In this frequency range (100--217 GHz) the CMB is the dominant
component, except for the Galactic plane region. Our simulated
skies contain a mixture of CMB, synchrotron and thermal emission,
described below. The dust emission templates have been obtained by
extrapolating the maps by Schlegel et al. (1998), which combine
IRAS and DIRBE data, assuming a grey-body spectrum (expressed in antenna
temperature), 
\begin{equation}
\label{dustscaling}
f_{dust}(\nu)\propto {\tilde{\nu}^{\beta+1}\over e^{\tilde{\nu}}-1}\ ,\
\tilde{\nu}={h\nu\over kT_{dust}}
\end{equation}
with uniform temperature $T_{dust}=18\,$K and emissivity $\beta=2$. 

Unfortunately, no maps of free-free emission are available at the
moment, although an all-sky map of $H_{\alpha}$ emission (which is
known to be a good tracer of free-free emission) will be available
in a couple of years (Reynolds \& Haffner, 2000). On the other
hand, at $\lsim 100\,$GHz free-free  emission may exceed the dust
emission component. In order to simulate the free-free
contribution we assume, somewhat arbitrarily, that it is perfectly
correlated with the dust itself, i.e. that it has the same spatial
distribution. Its antenna temperature scales with frequency as
$T_{A,ff}\propto \nu^{-\beta_{ff}}$, with $\beta_{ff}=-2.1$. The
relative amplitude of dust and free-free emission is assumed to be
a factor of 3 at 100 GHz (De Zotti et al. 1999). Thus, for the
{\ica} analysis, free-free and dust emission effectively behave as
a single component, referred to, in the following, as the
``thermal'' component, having a spectrum described by:
\begin{equation}
\label{thermalscaling} f_{thermal}(\nu )={1\over 3}\left({\nu
\over 100{\rm GHz}}\right)^{\beta_{ff}}+ {f_{dust}(\nu )\over
f_{dust}(100 {\rm GHz})}\ .
\end{equation}

\begin{figure*}
\begin{center}
\epsfig{file=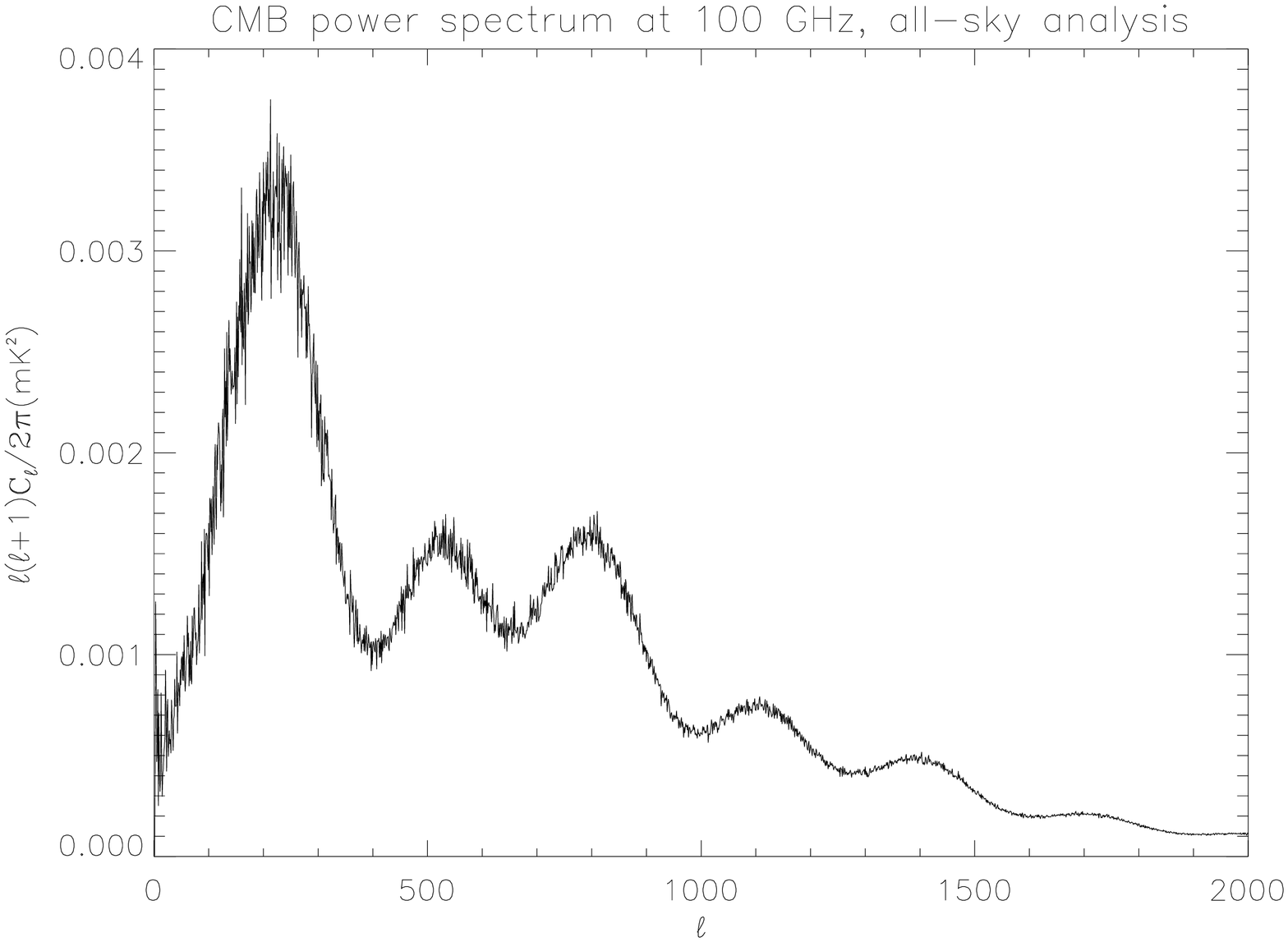,height=3.in,width=2.in,angle=90}
\hskip .2in
\epsfig{file=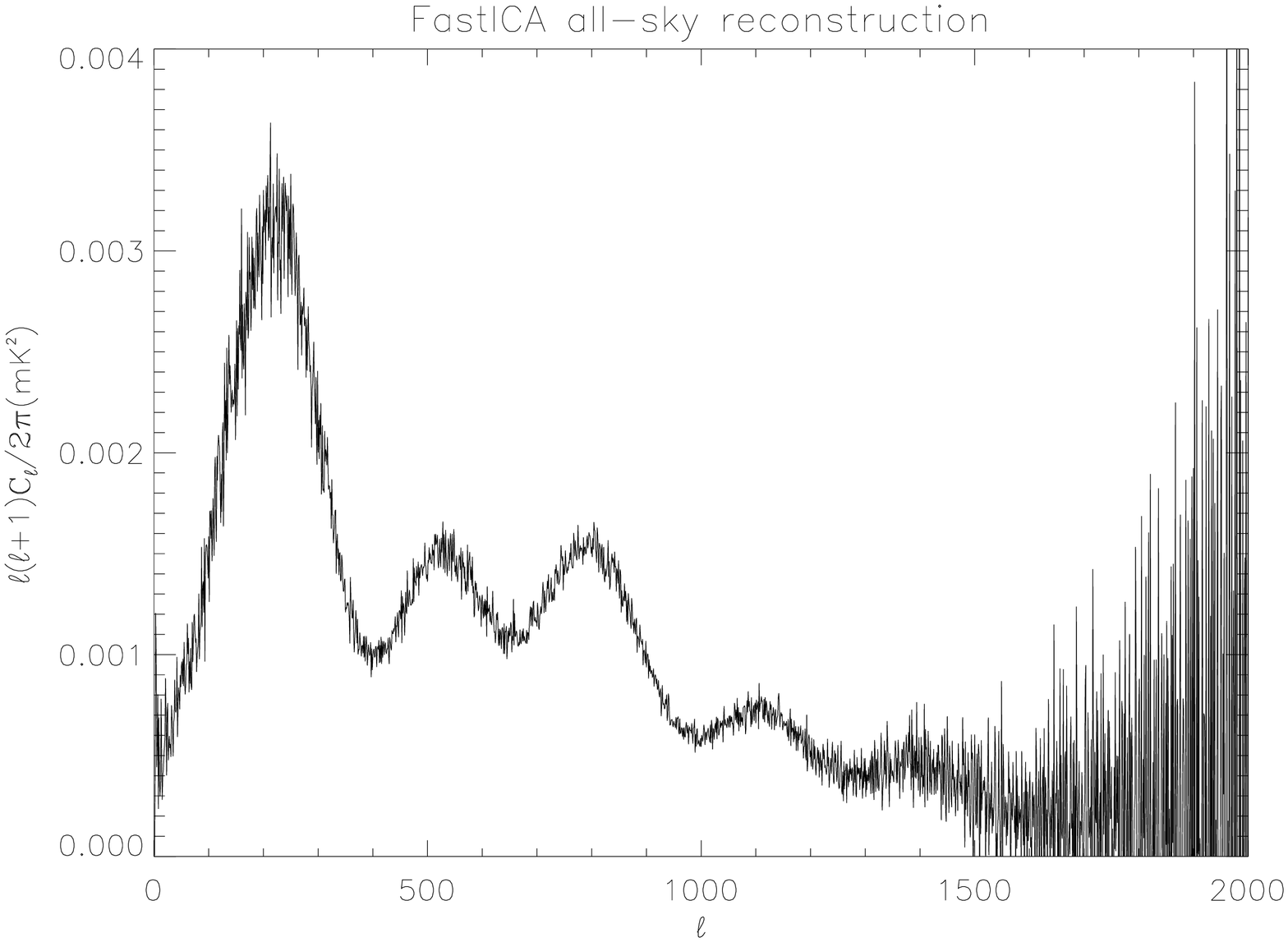,height=3.in,width=2.in,angle=90}
\vskip .3in
\epsfig{file=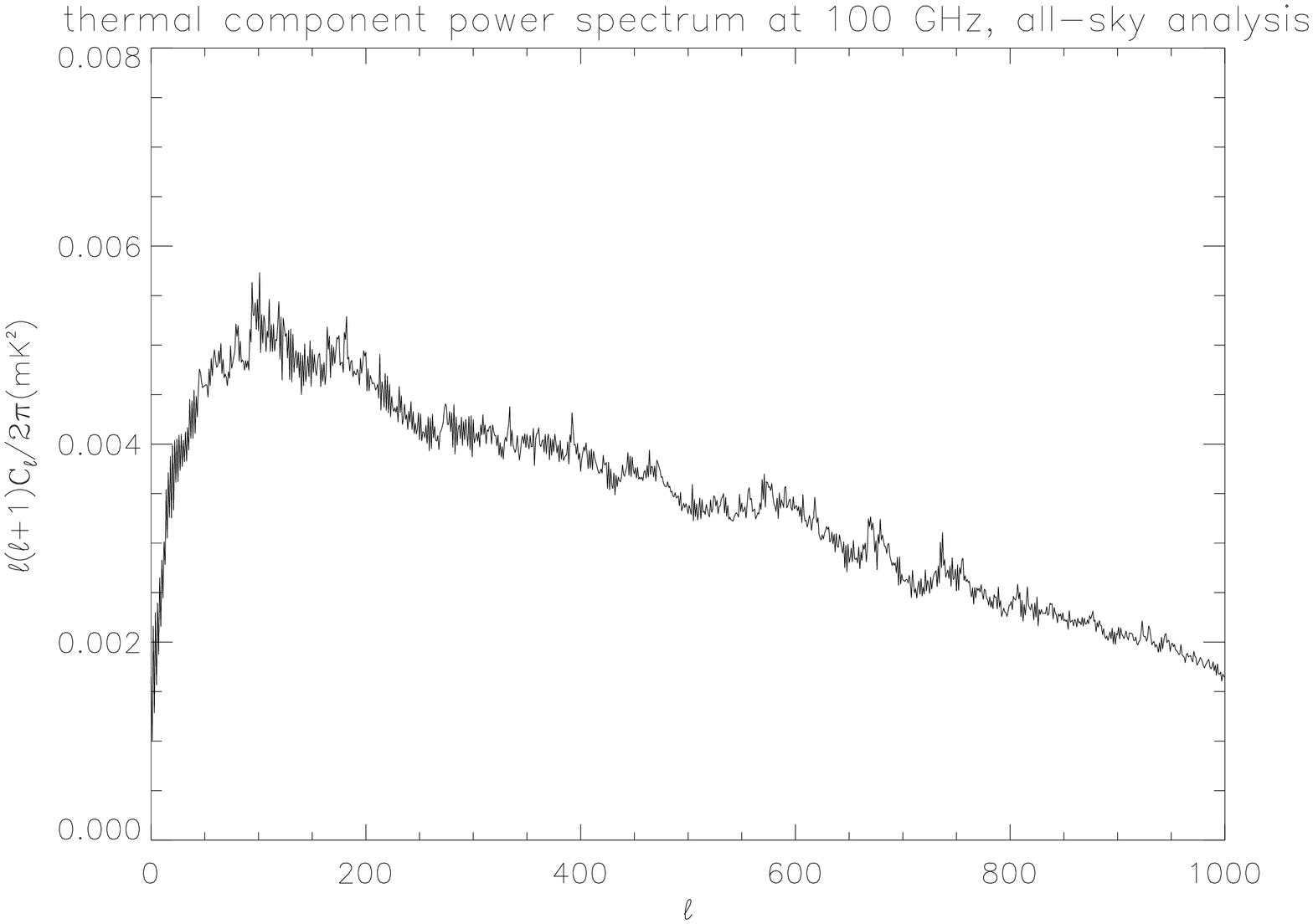,height=3.in,width=2.in,angle=90}
\hskip .2in
\epsfig{file=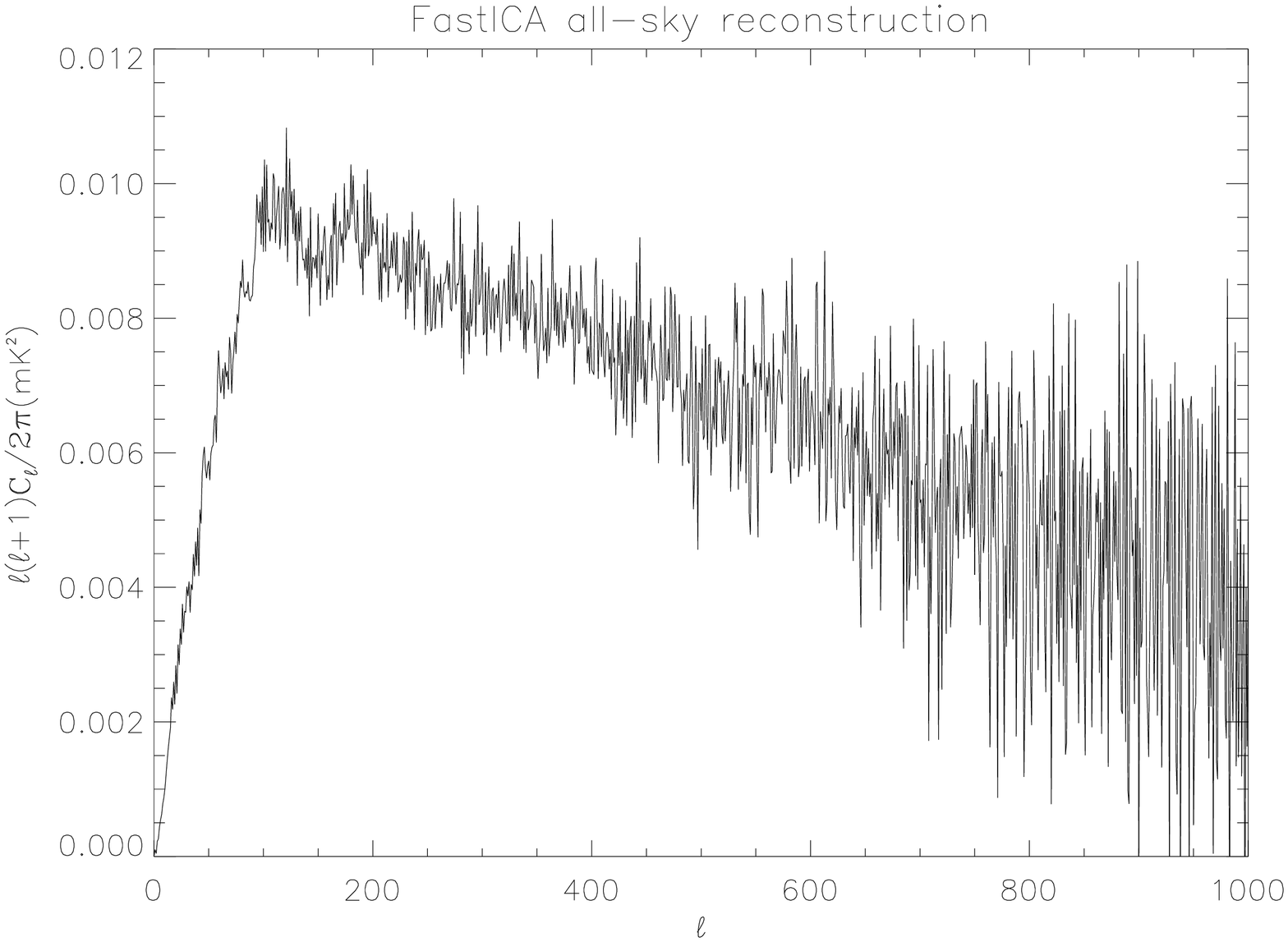,height=3.in,width=2.in,angle=90}
\vskip .3in
\epsfig{file=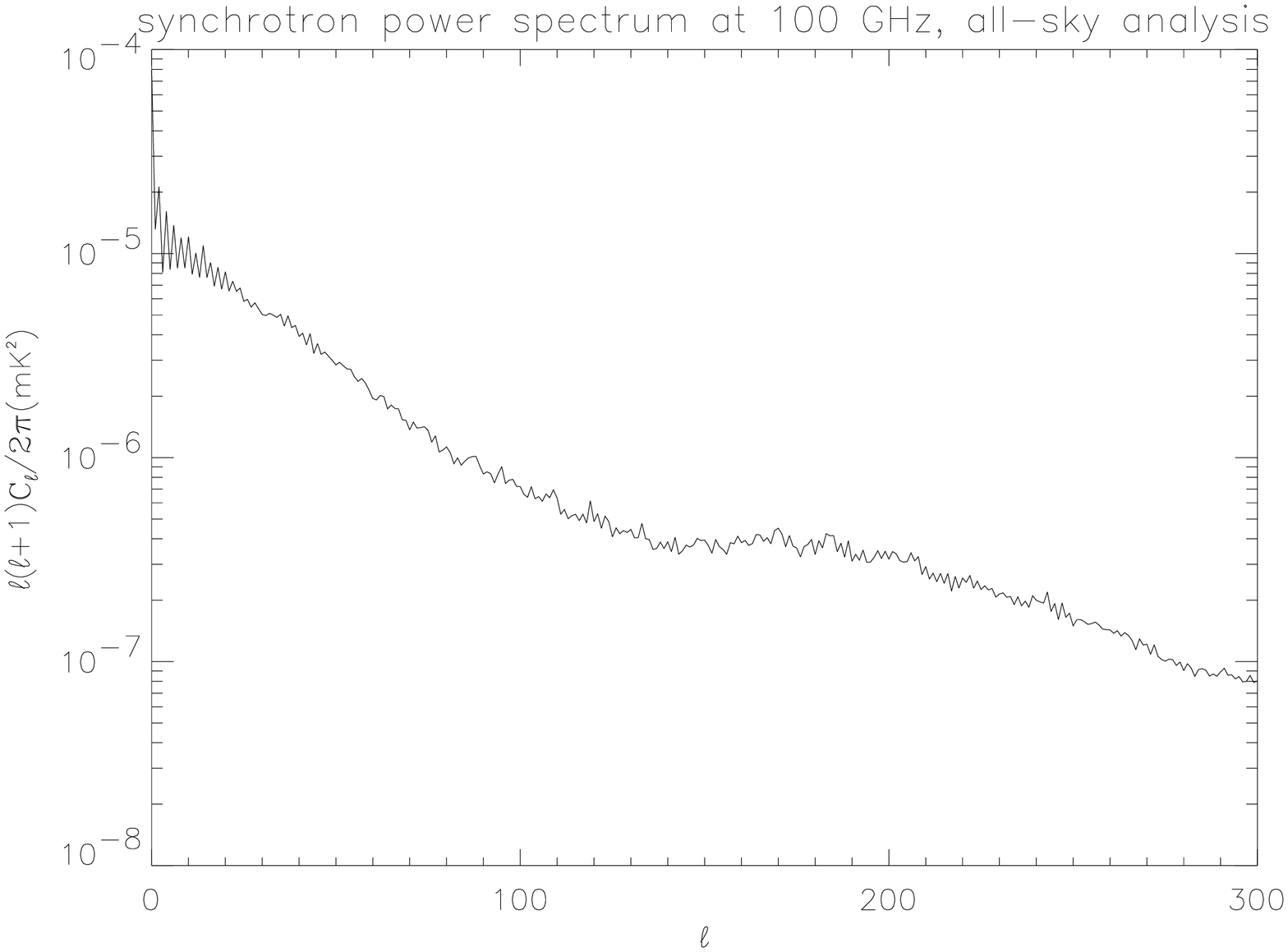,height=3.in,width=2.in,angle=90}
\hskip .2in
\epsfig{file=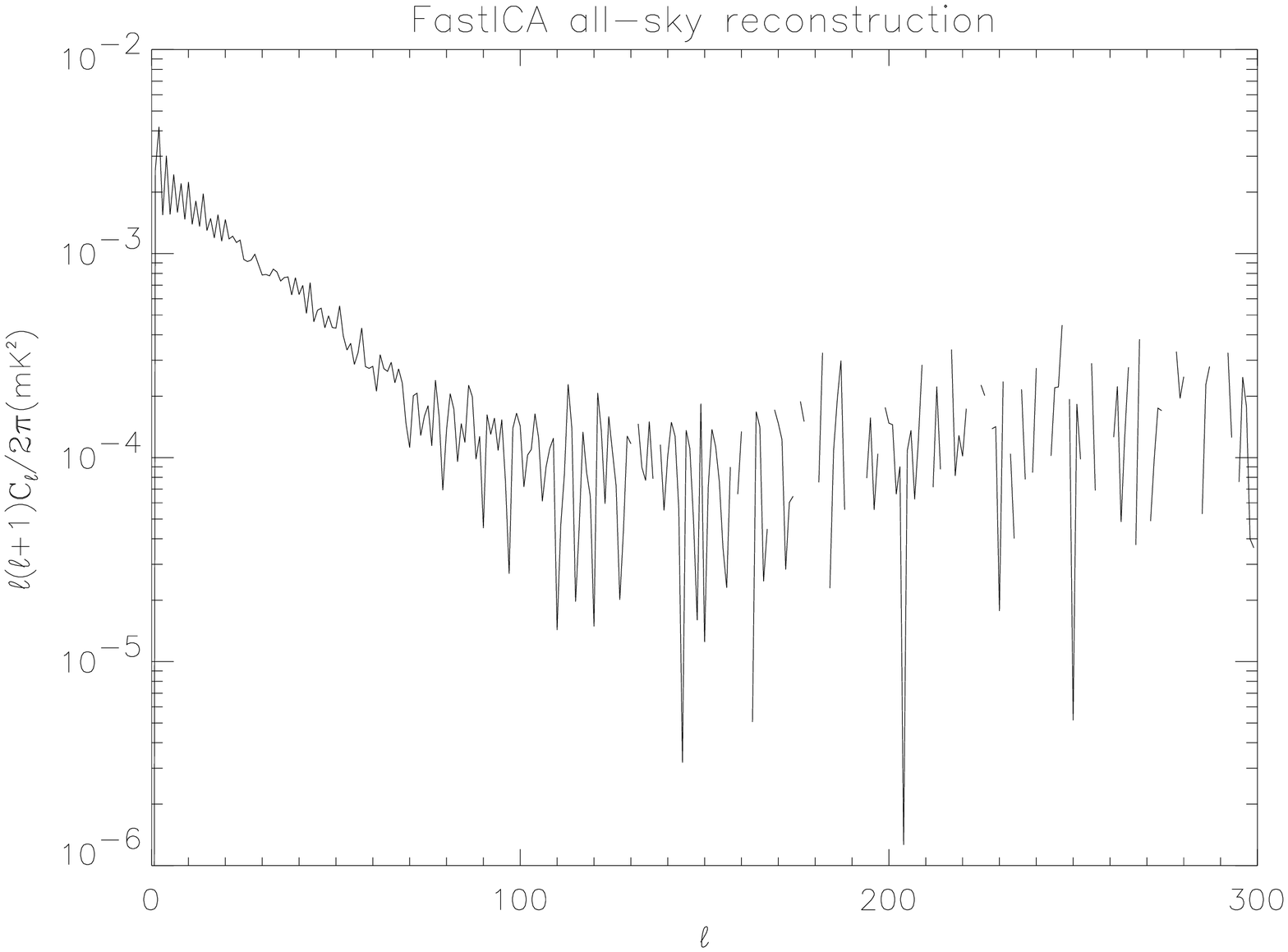,height=3.in,width=2.in,angle=90}
\vskip .3in \caption{Input (left) and output (right) angular power
spectra of the CMB, of the thermal component, and of synchotron
emission for case 2, at 100 GHz.} \label{clcaso2}
\end{center}
\end{figure*}

The Galactic signals are well below the noise at medium and high
Galactic latitudes so that no reconstruction is possible there at
these frequencies. Therefore we have restricted the analysis to
all-sky maps. The smoothing of the noisy sky maps has been
performed with an FWHM of 10$'$, and we have used $N_{\rm
side}$=1024 (3.5$'$ pixels).
\begin{figure*}
\begin{center}
\epsfig{file=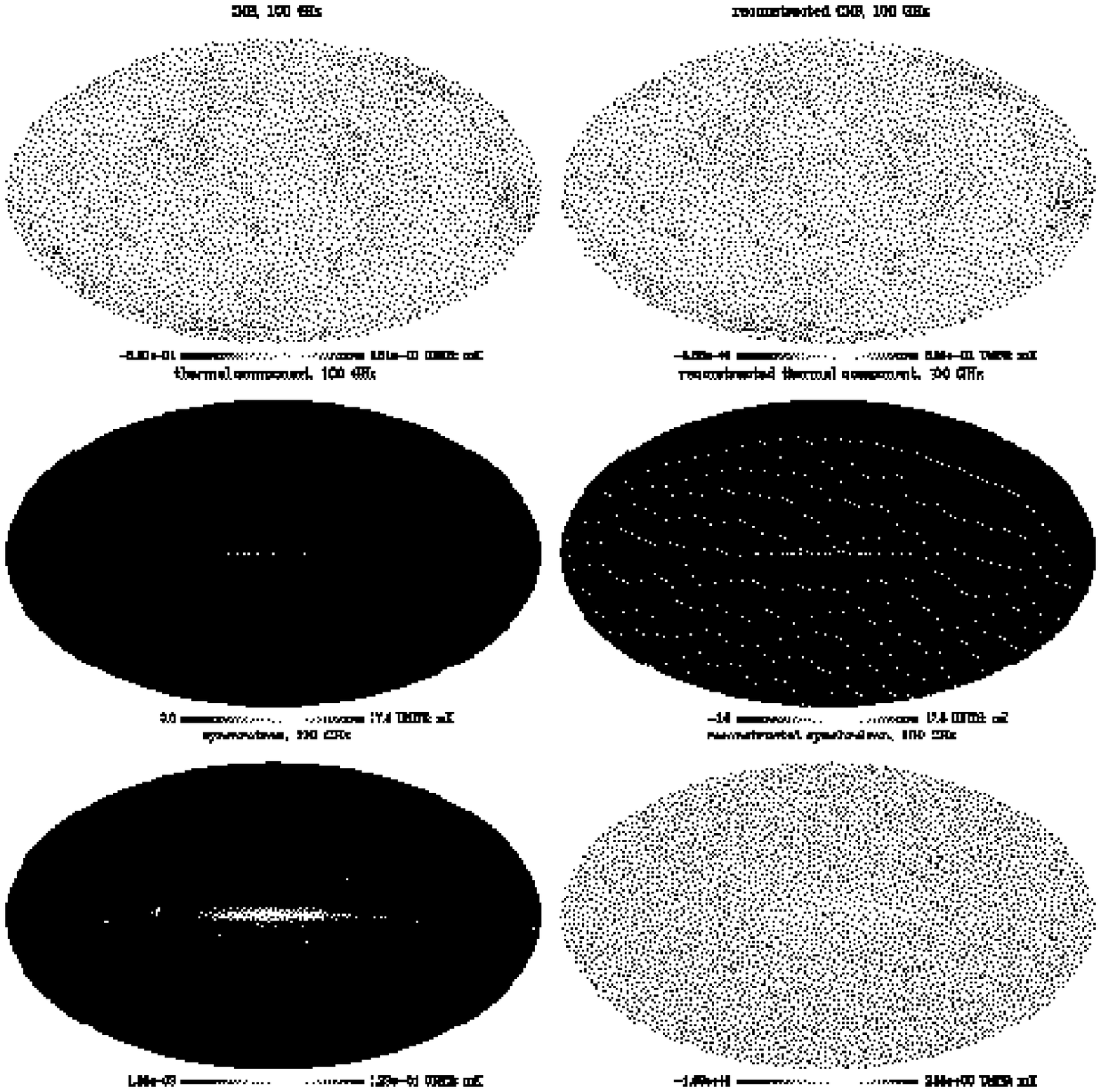,width=7.in,height=9.in}
\vskip -1.5in
\caption{Input (left) and output (right) maps for case 2, at 100
GHz.} \label{caso2}
\end{center}
\end{figure*}
Figures \ref{caso2} and \ref{clcaso2} show the input (left) and
output (right) maps and power spectra, respectively, at 100 GHz.

The CMB power spectrum is reconstructed up to $\ell\simeq 1500$
with the correct normalization. On the contrary, thermal emission
and synchrotron power spectra are much more affected by noise. The
power spectrum of thermal emission is quite well reconstructed up
to $\ell\simeq 600$, but its normalization is overestimated by a
factor $\simeq 1.8$. The shape of the synchrotron power spectrum
is recovered up to $\ell \simeq 100$, but its normalization was
lost.

The reason of the bad quality of the synchrotron reconstruction
resides in the weakness of this component at these frequencies,
with respect to noise, and in the poor statistics contained 
in the synchrotron template we use, as we already stressed. 

However, it is remarkable that {\ica} could separate, to some extent,
thermal and synchrotron emissions, despite their low
signal-to-noise ratio and the correlation in their spatial
distributions (both are maximum on the Galactic plane). The latter
point illustrates the fact that the statistical independence
required for {\ica} to work is not so much related to the spatial
pattern of the components to be separated but, rather, to their
distribution functions. The templates we have adopted for thermal
and synchrotron emissions do have quite different distribution
functions, as illustrated by Table~\ref{momenti}, giving their
skewness and kurtosis. This makes {\ica} able to perform some
separation of them.

\begin{figure*}
\begin{center}
\epsfig{file=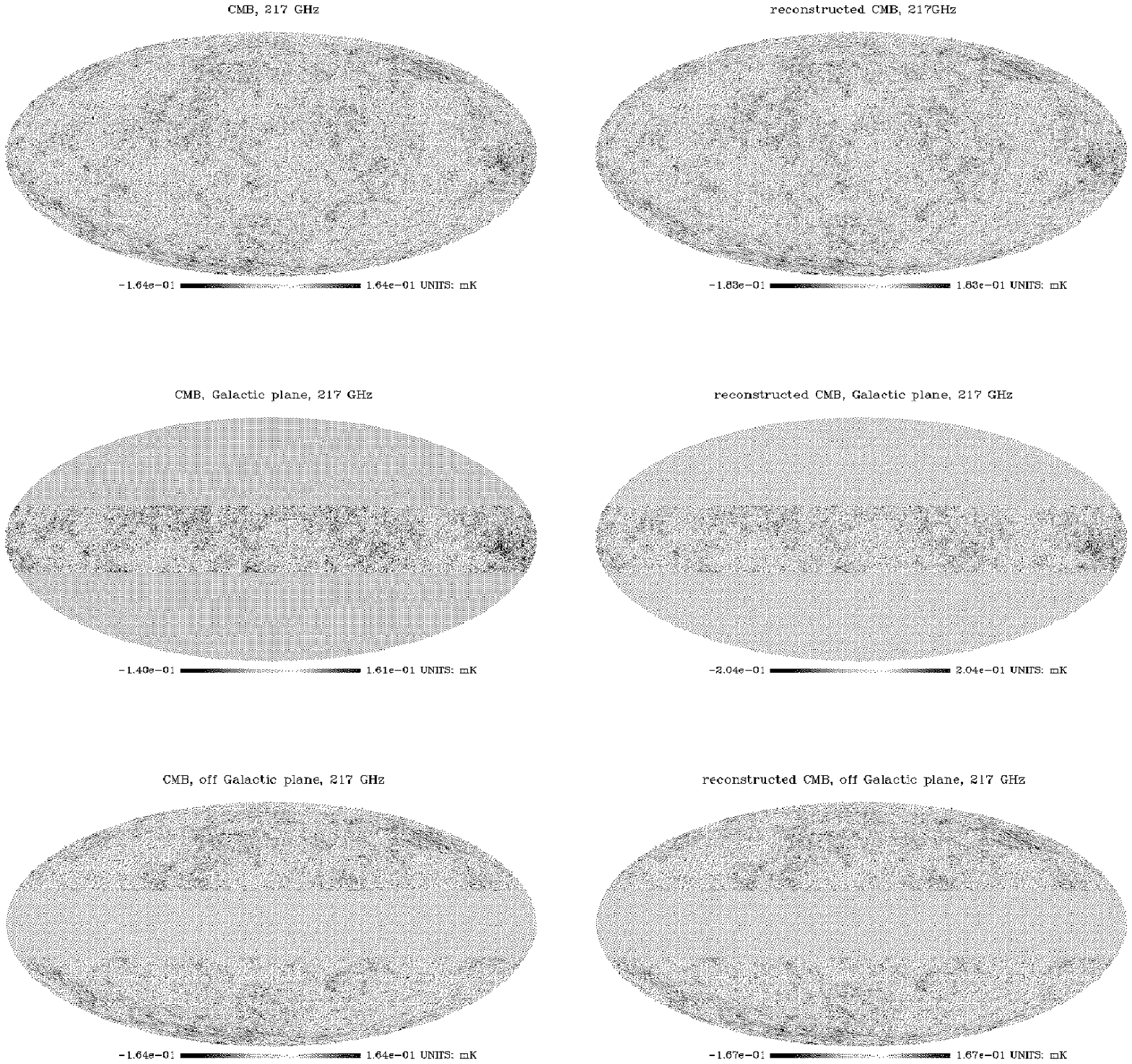,width=7.in,height=9.in}
\vskip -1.5in
\caption{Input (left) and output (right) CMB maps for case 3.}
\label{CMBcaso3}
\end{center}
\end{figure*}

\begin{table}
\begin{center}
\caption{Skewness and kurtosis of CMB, thermal component and
synchrotron templates.} \label{momenti}
\begin{tabular}{lccc}
\hline
\ $Component$ & Skewness & Kurtosis \\
\hline
\ CMB  & 0.014 &  0.087 \\
\ Thermal & 51.7692  &  5098.16  \\
\ Synchrotron & 7.717  &  80.693 \\
\hline
\end{tabular}
\end{center}
\end{table}
As we can see from Table~\ref{tablescalings}, frequency scalings
are recovered very well, at the percent level, for the CMB and
thermal components, while the error is considerably larger for
synchrotron, which is the weakest component in this case.

\subsection{Case 3}
For this analysis, we selected two channels, 217 and 353 GHz, and
two components, CMB and thermal emission which, at these
frequencies, is dominated by dust. Maps with  $N_{\rm side}$=1024
have been smoothed with a FWHM of 5$'$. As in case 1, we have made
separate analyses for the whole sky, and for the high and low
Galactic latitude regions. The average signal-to-noise ratio for
the dust is 190 over the whole sky, 246 for $|b|<20^\circ$, and
6.6 at high Galactic latitude; for the CMB it is $\sim 5.7$.

\begin{figure*}
\begin{center}
\epsfig{file=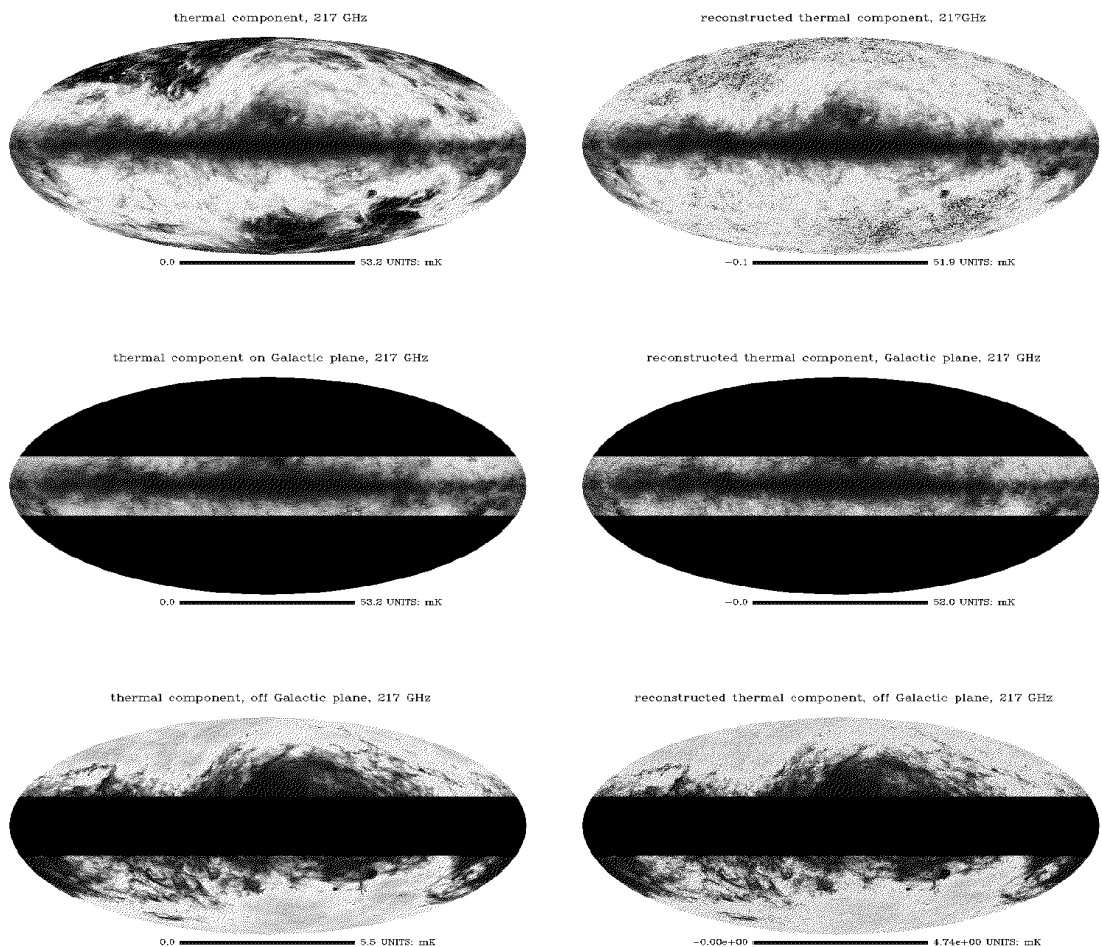,width=7.in,height=9.in}
\vskip -1.5in
\caption{Input (left) and output (right) thermal emission maps for
case 3, at 217 GHz.} \label{thecaso3}
\end{center}
\end{figure*}

\begin{figure*}
\begin{center}
\epsfig{file=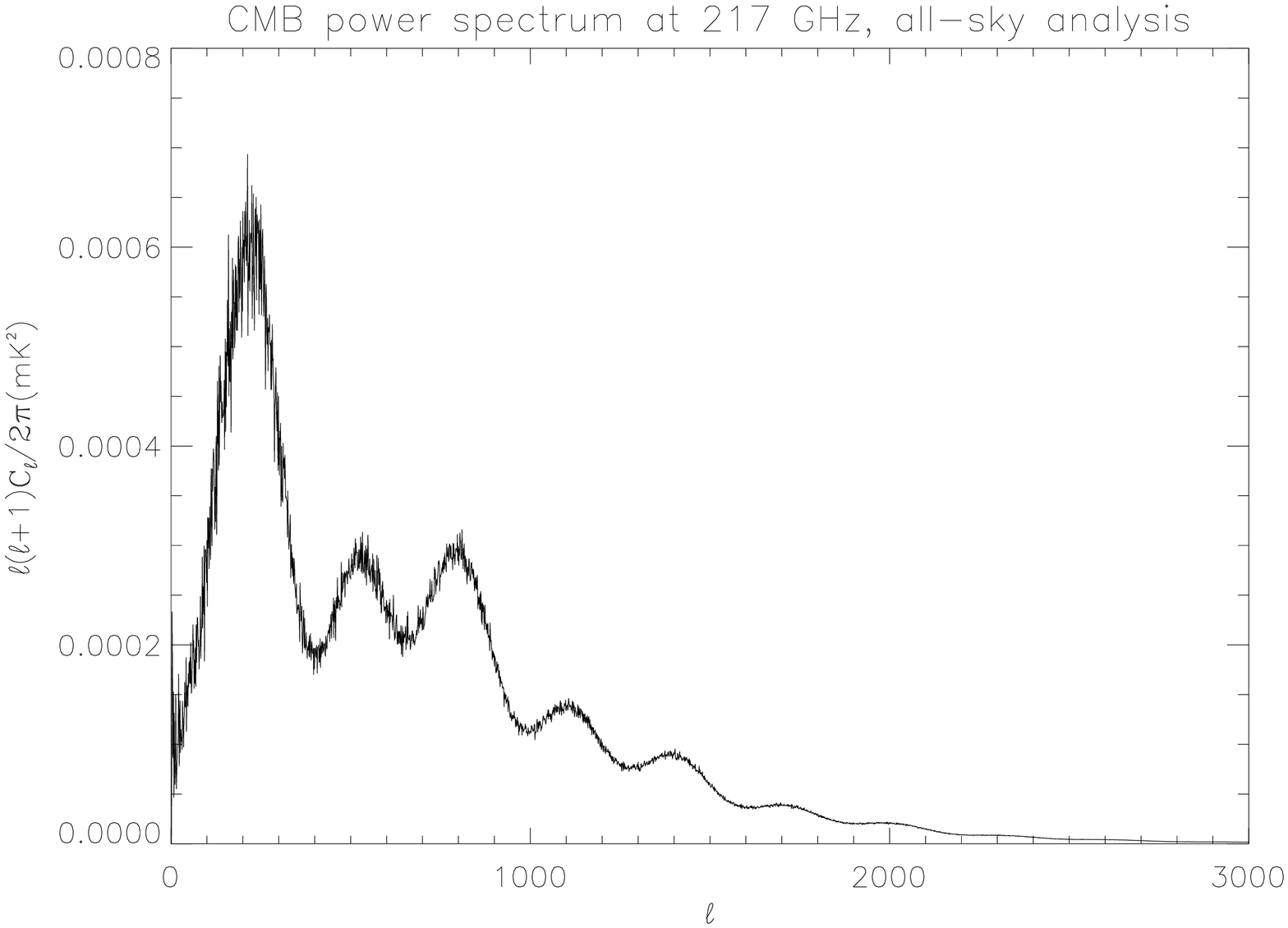,height=3.in,width=2.in,angle=90}
\hskip .2in
\epsfig{file=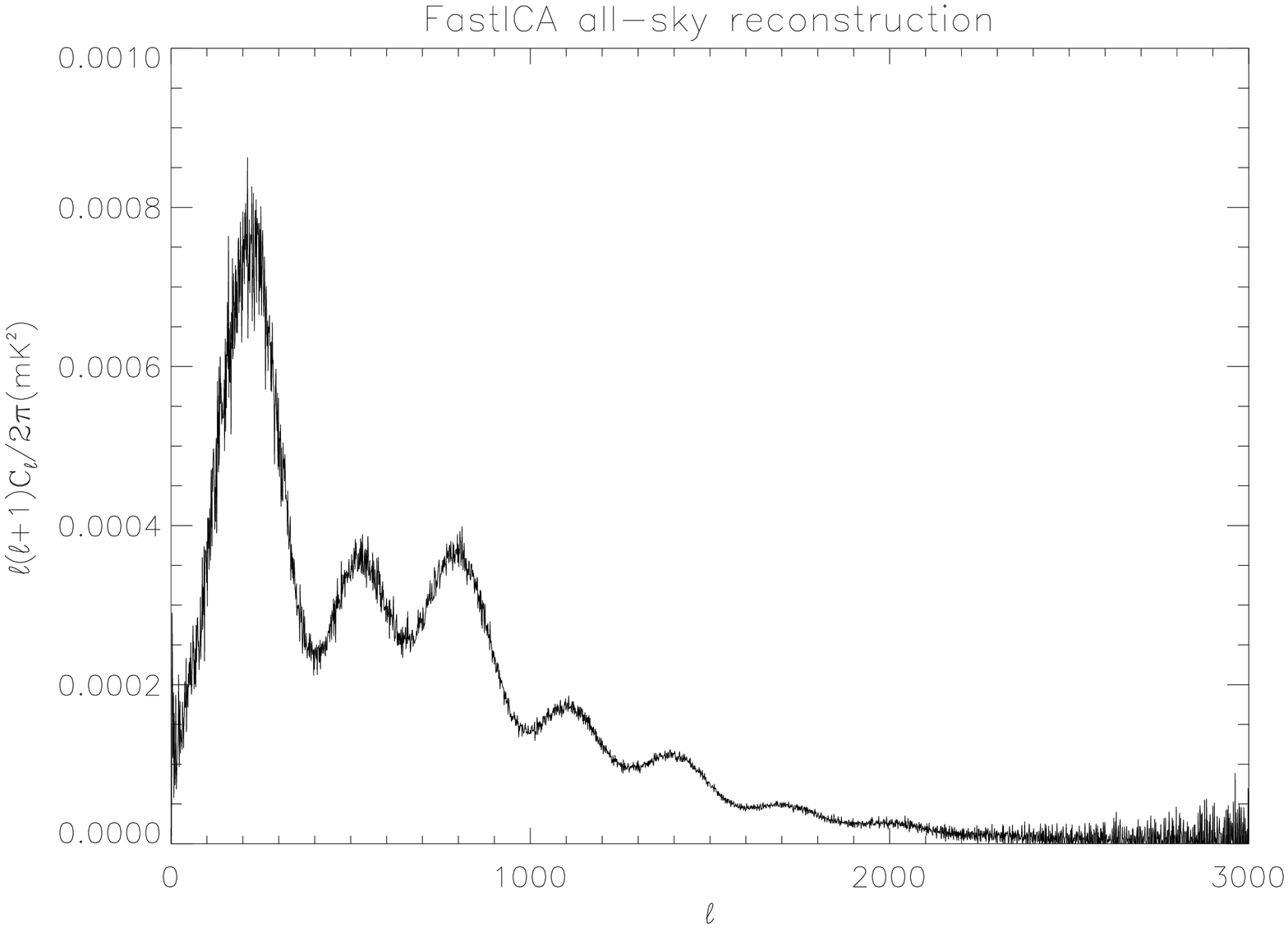,height=3.in,width=2.in,angle=90}
\vskip .3in
\epsfig{file=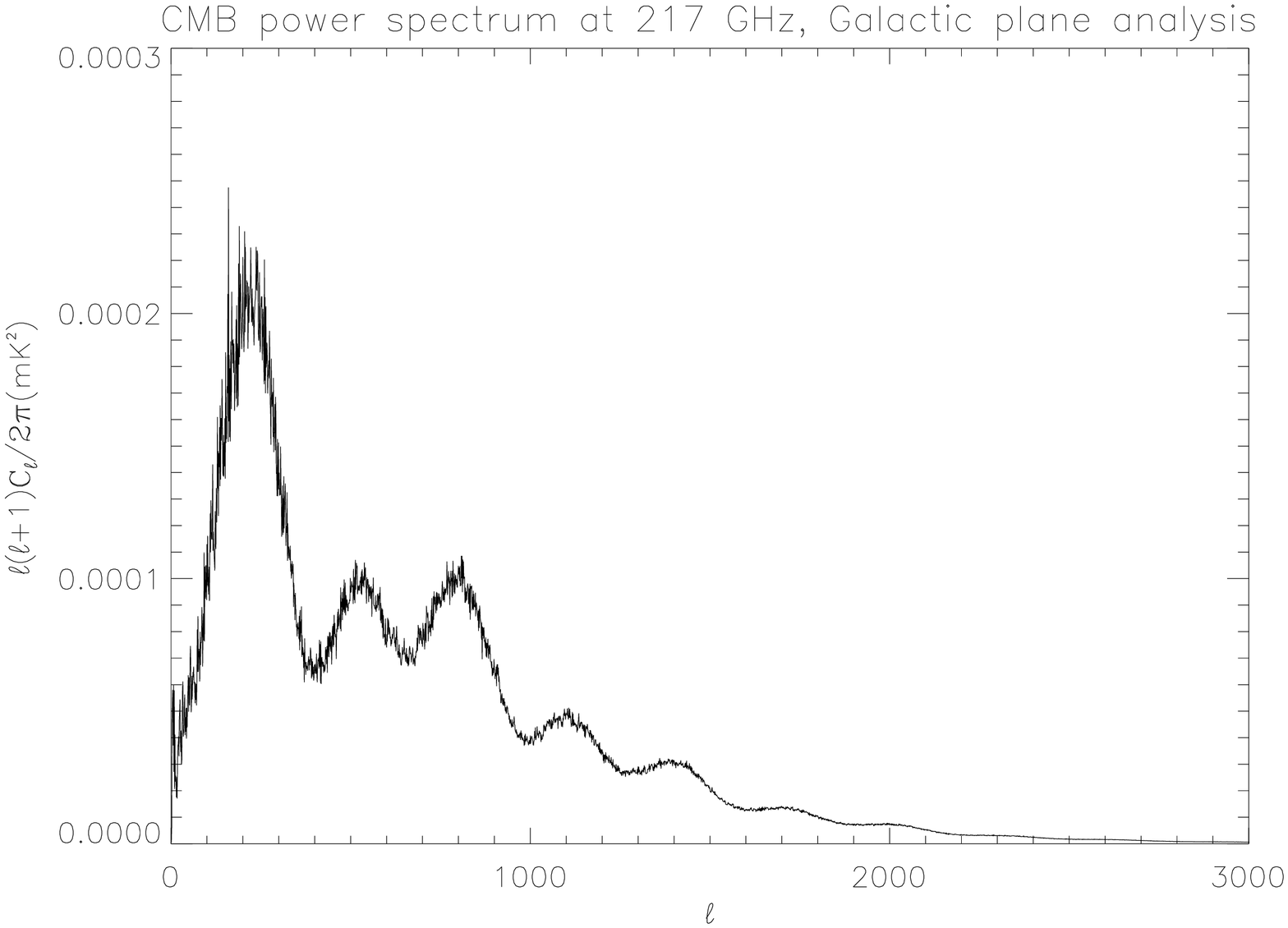,height=3.in,width=2.in,angle=90}
\hskip .2in
\epsfig{file=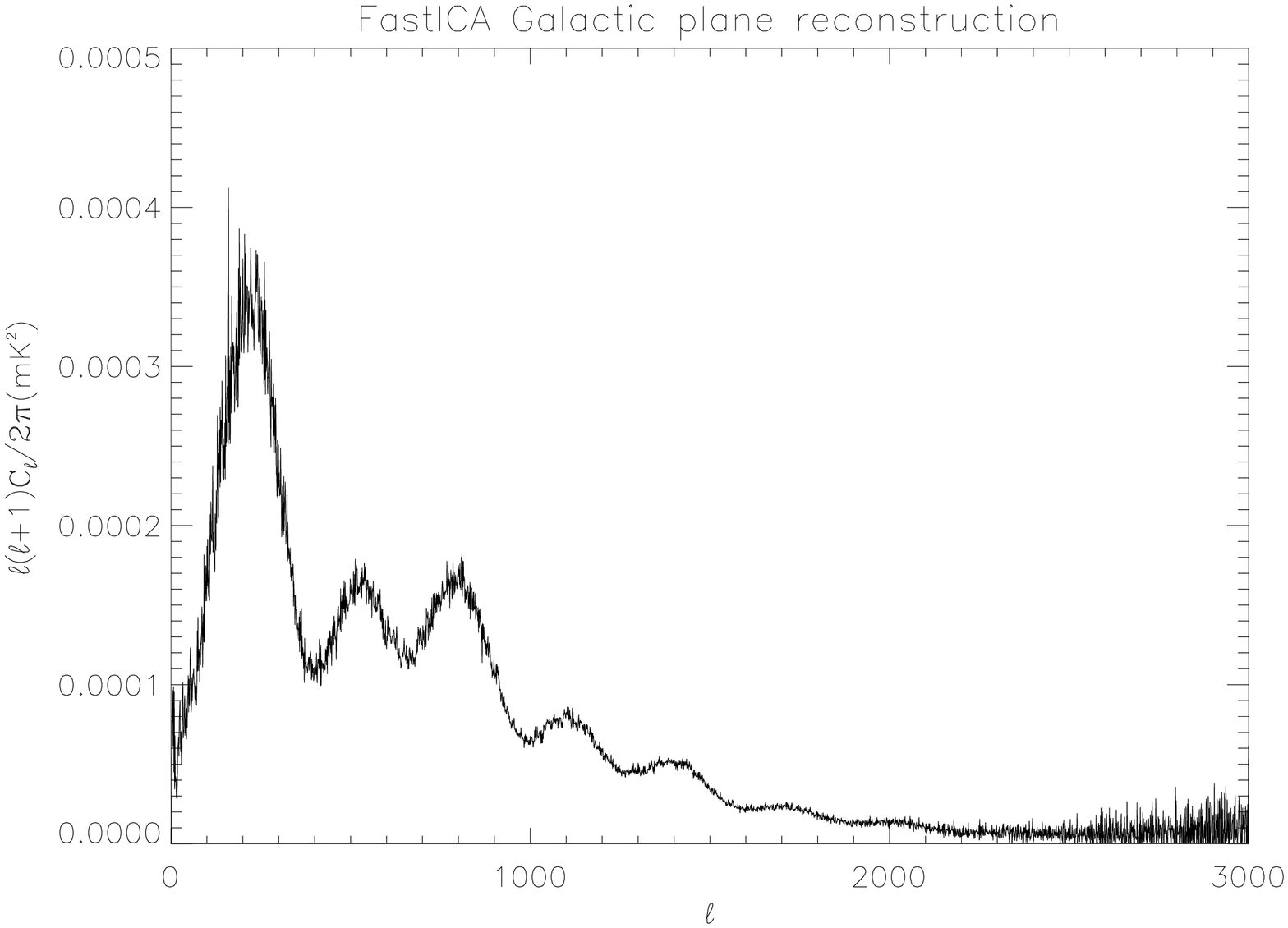,height=3.in,width=2.in,angle=90}
\vskip .3in
\epsfig{file=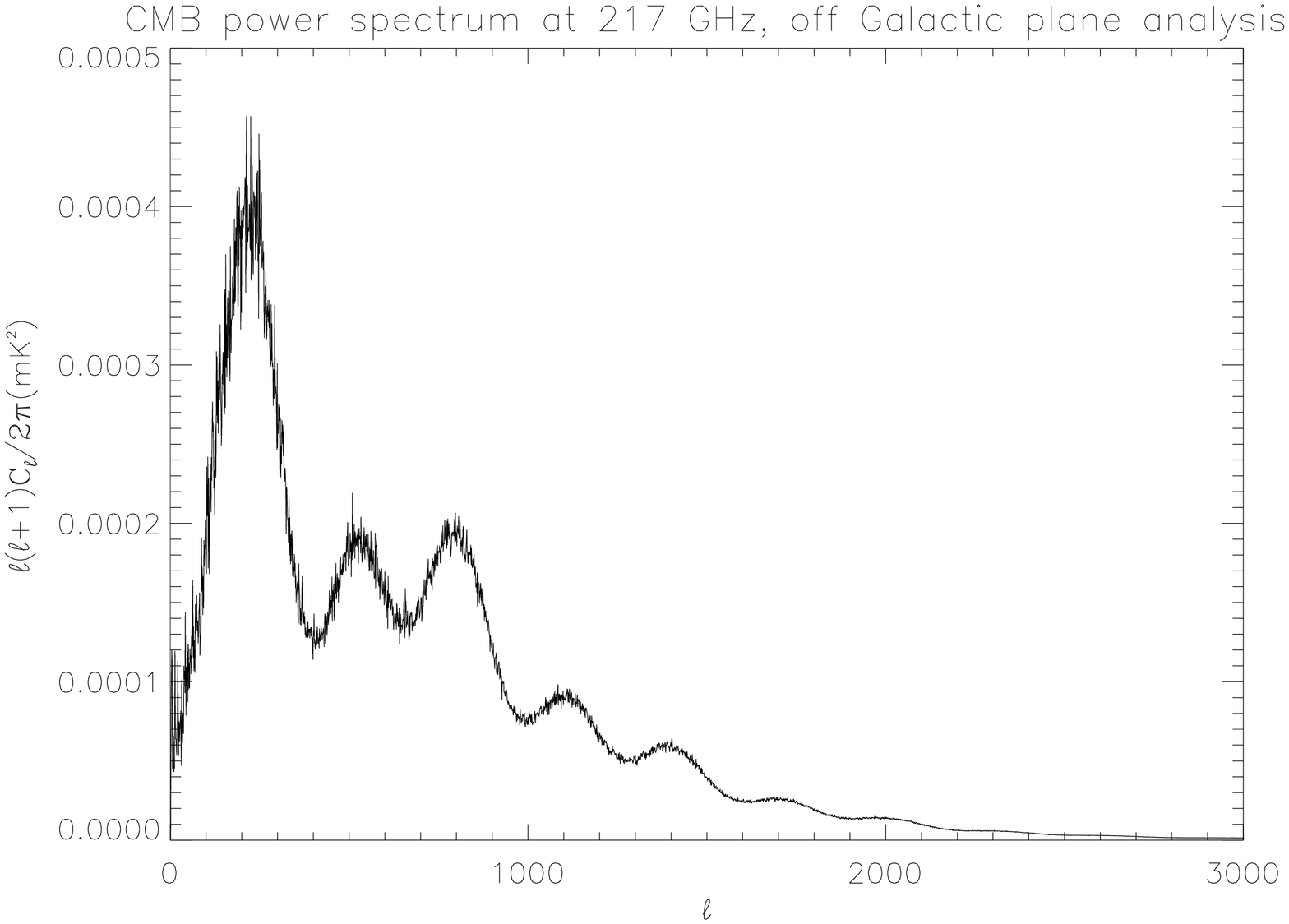,height=3.in,width=2.in,angle=90}
\hskip .2in
\epsfig{file=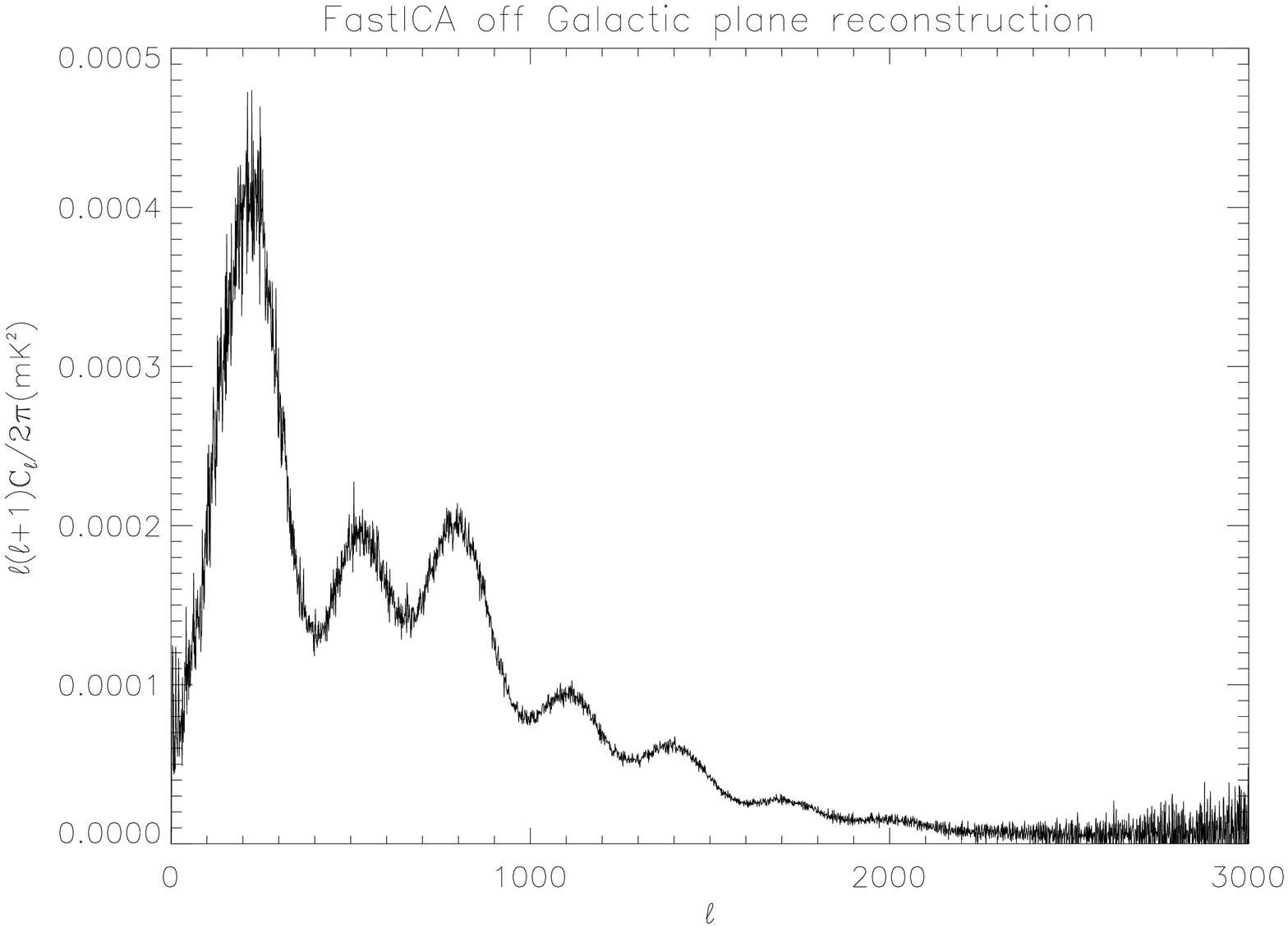,height=3.in,width=2.in,angle=90}
\vskip .3in \caption{Input (left) and output (right) CMB angular
power spectra for case 3, at 217 GHz. From top to bottom: all-sky,
low and high Galactic latitude results, respectively.}
\label{cl_CMBcaso3}
\end{center}
\end{figure*}

\begin{figure*}
\begin{center}
\epsfig{file=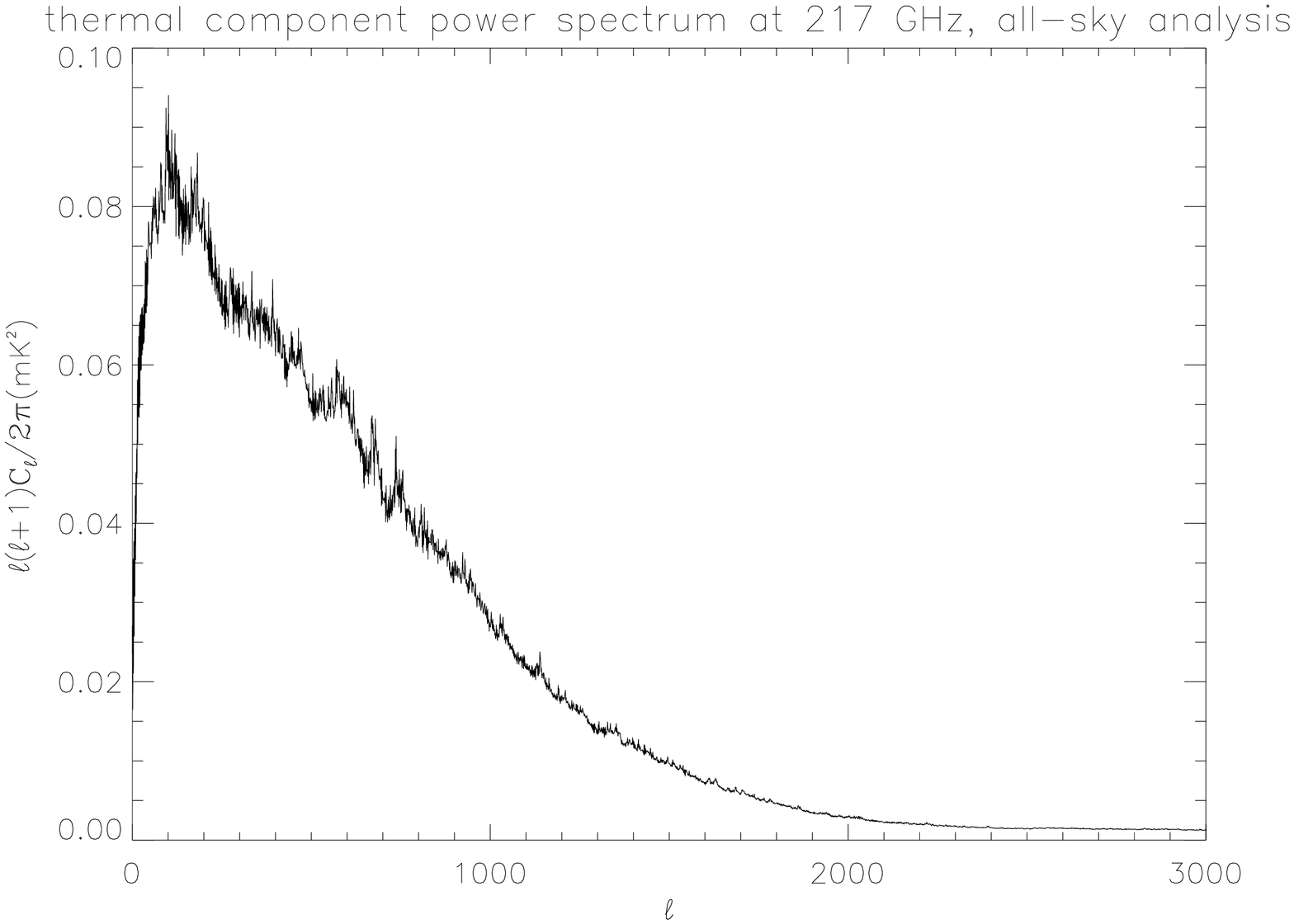,height=3.in,width=2.in,angle=90}
\hskip .2in
\epsfig{file=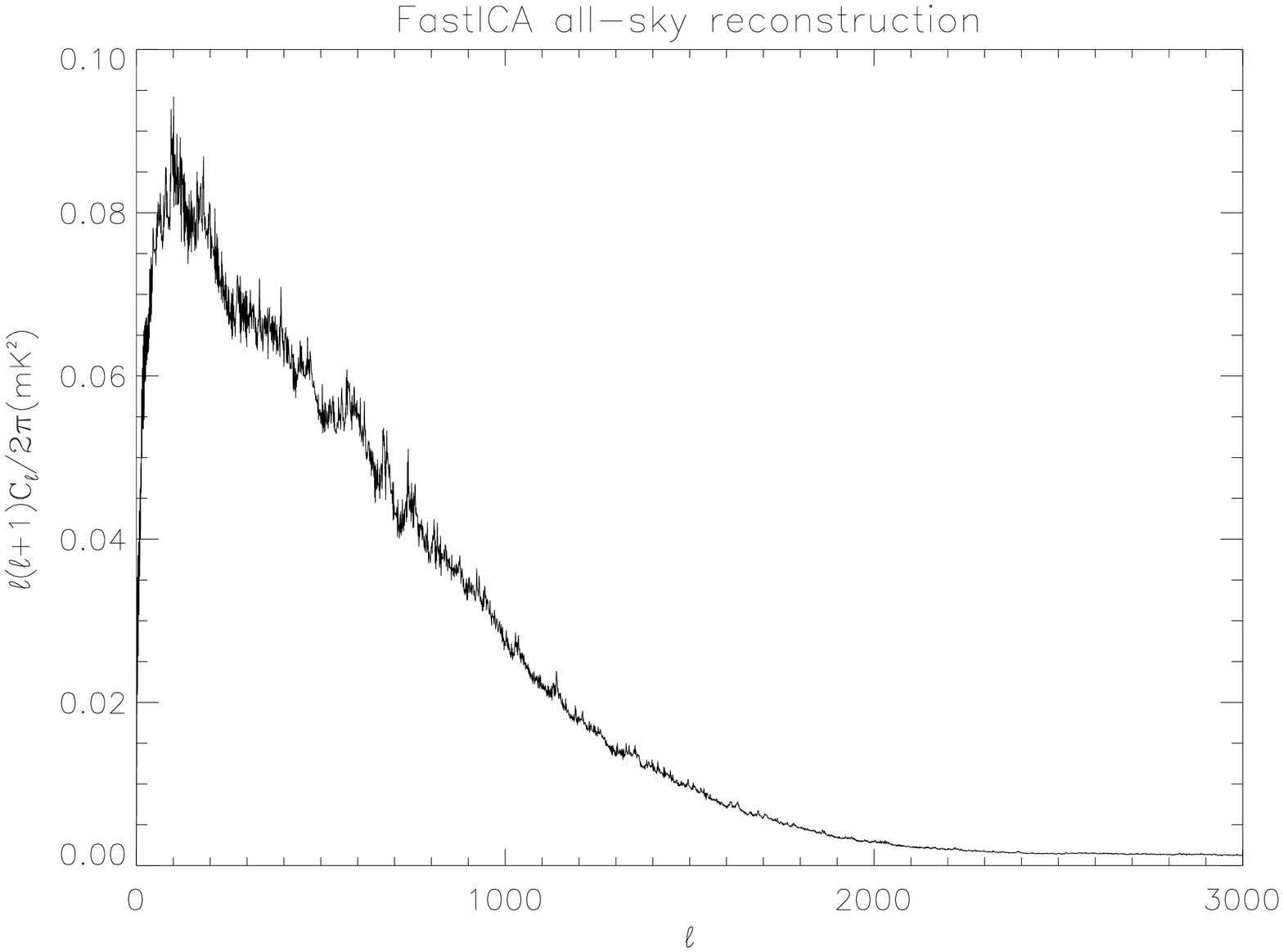,height=3.in,width=2.in,angle=90}
\vskip .3in
\epsfig{file=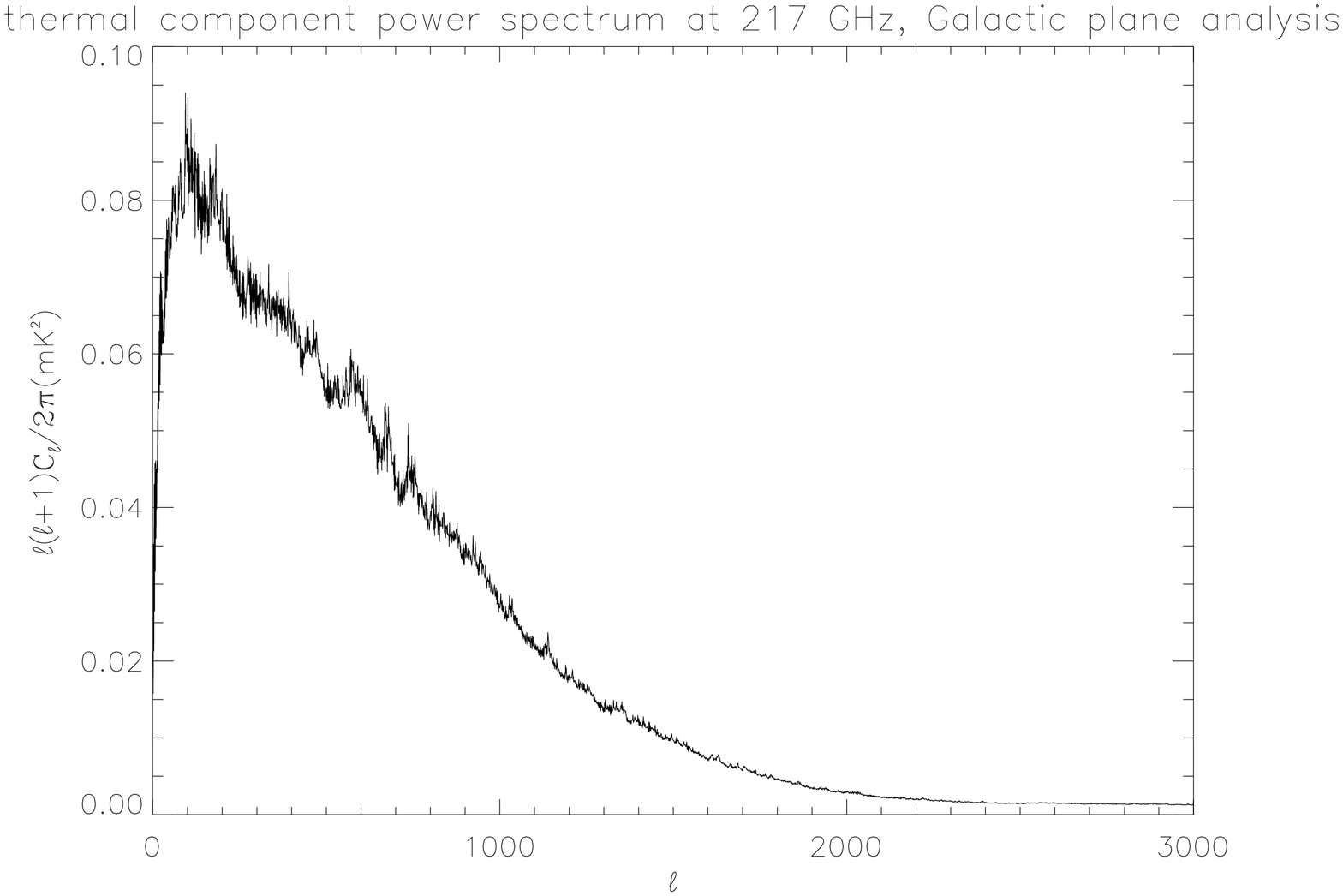,height=3.in,width=2.in,angle=90}
\hskip .2in
\epsfig{file=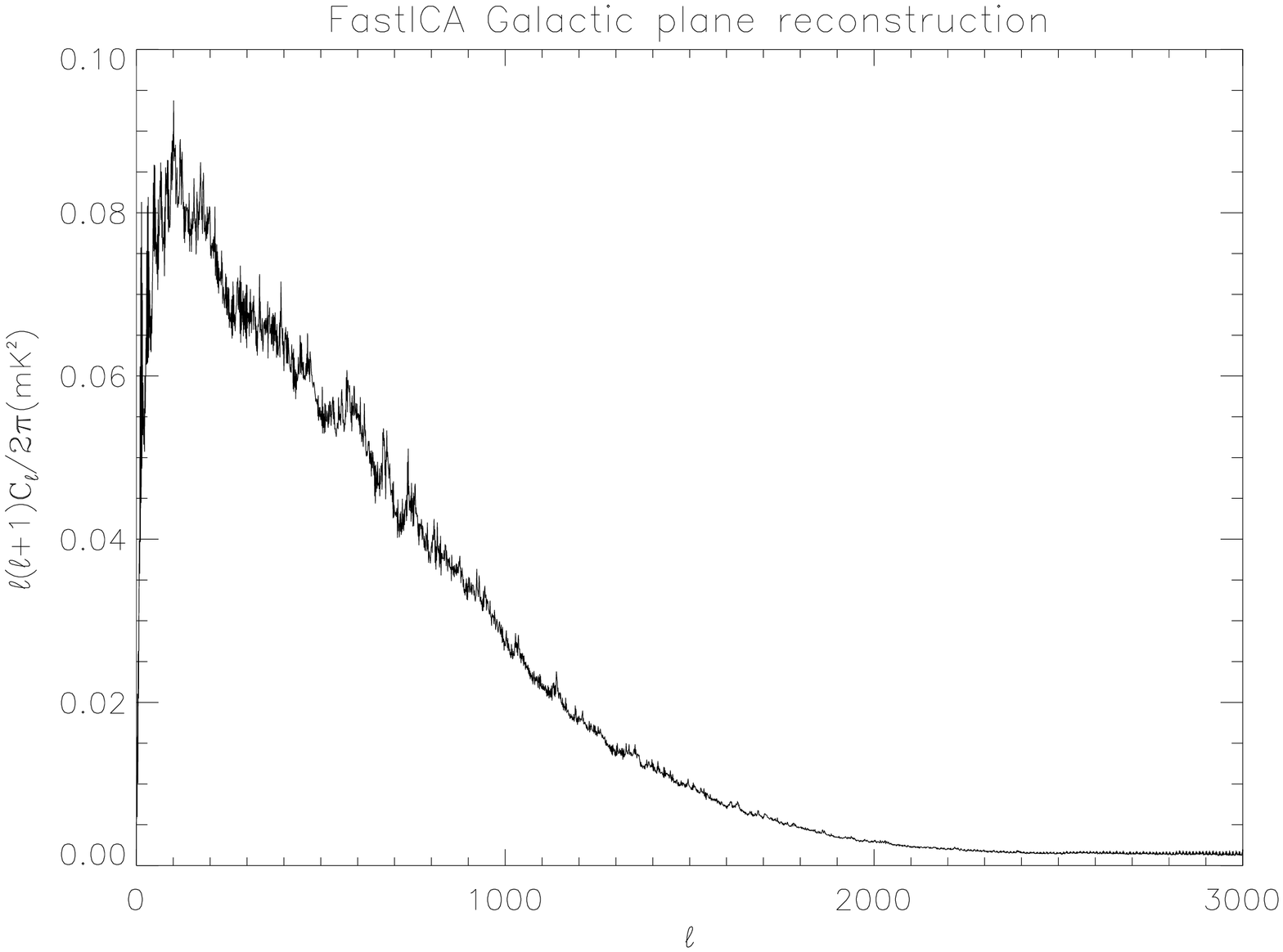,height=3.in,width=2.in,angle=90}
\vskip .3in
\epsfig{file=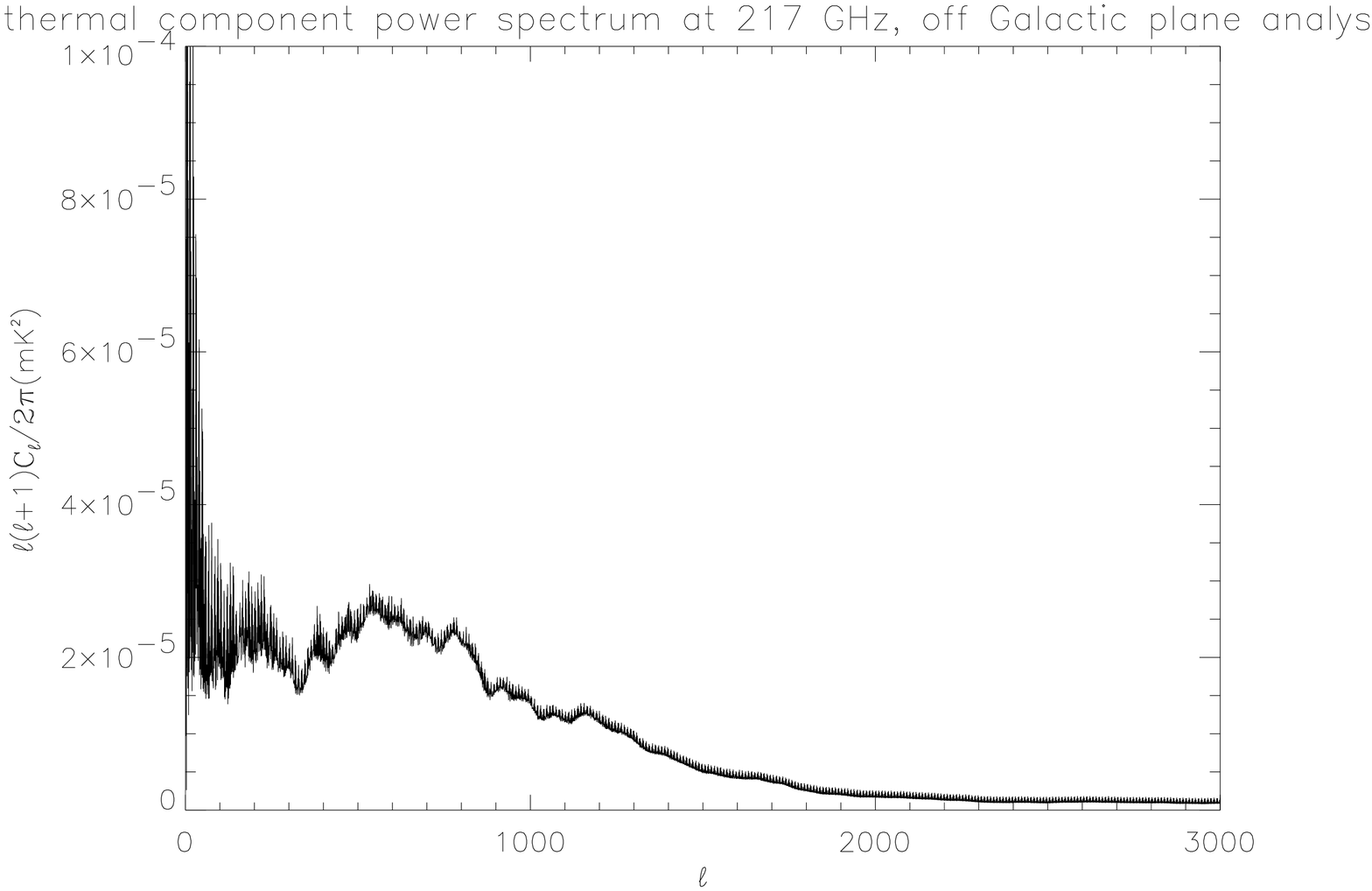,height=3.in,width=2.in,angle=90}
\hskip .2in
\epsfig{file=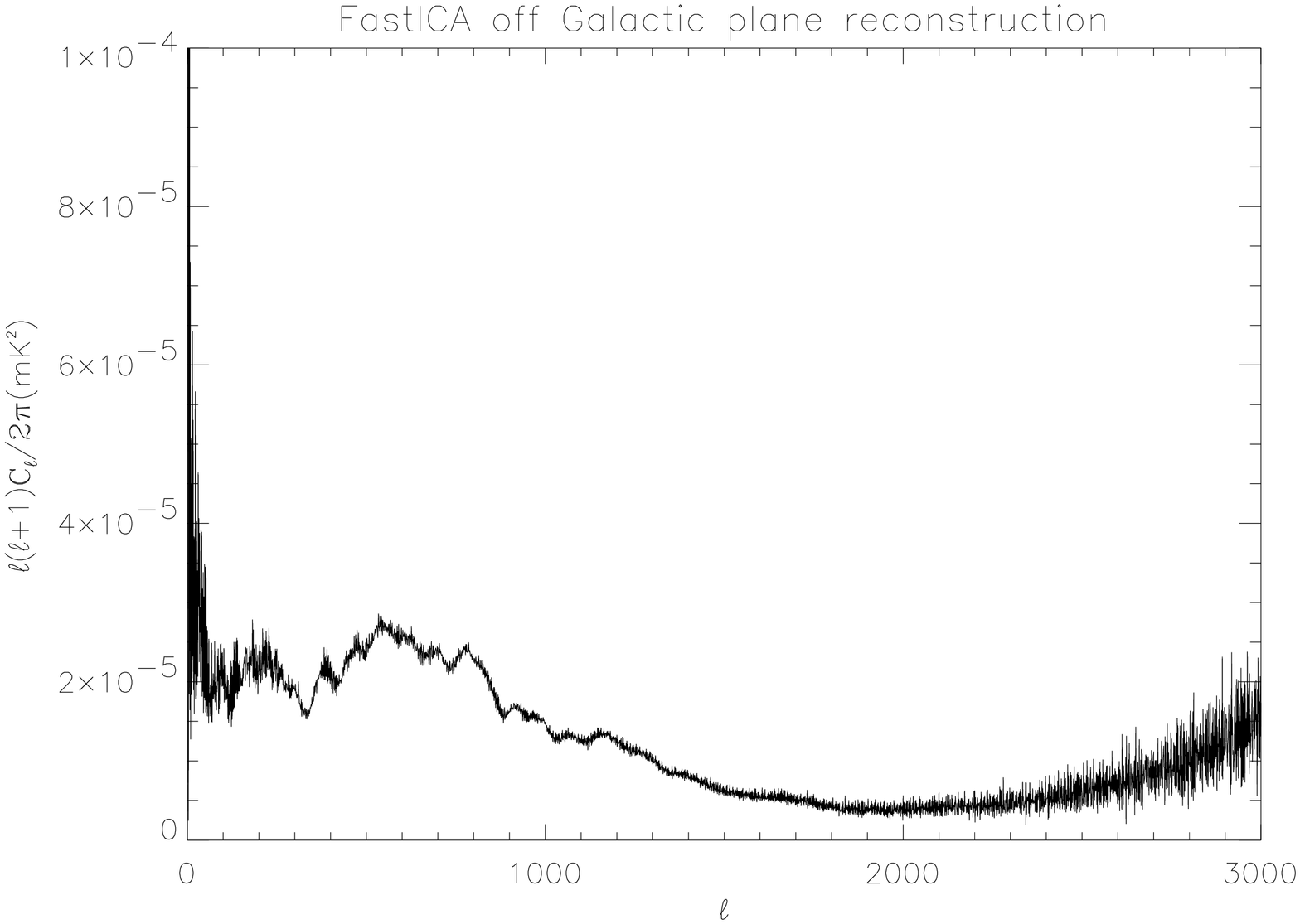,height=3.in,width=2.in,angle=90}
\vskip .3in \caption{Input (left) and output (right) angular power
spectra of the thermal component for case 3 at 217 GHz. From top
to bottom: all-sky, low, and high Galactic latitude results,
respectively.} \label{cl_thecaso3}
\end{center}
\end{figure*}

The results are reported in Figures.~\ref{CMBcaso3}, \ref{thecaso3},
\ref{cl_CMBcaso3}, and \ref{cl_thecaso3}, referring to 217 GHz.
The shapes of the power spectra of both signals are reconstructed
very well up to $\ell\simeq 2000$ in all cases shown.
The amplitude of the power spectrum of thermal emission is
accurately recovered in all cases, while in the case of the CMB
this happens only at high Galactic latitudes, where the two
components are comparable. As shown by Table~\ref{tablescalings},
the frequency scaling of thermal emission is recovered to a
fraction of 1\%, while for the CMB the error is much larger.

\section{Discussion and Conclusion}
\label{conclusions} We have presented a first application of a
new, computationally fast, technique for separation of
astrophysical components, based on Independent Component Analysis
concepts, {\ica}. The technique has been applied both to toy
models, useful to highlight some of its key properties, and to
simulated all-sky maps, with reference to the forthcoming {\sc
Planck} mission (Mandolesi et al. 1998; Puget et al. 1998).

We have achieved a considerable improvement over the previous
application of ICA-type algorithms to astrophysical component
separation (see Baccigalupi et al. 2000). First of all, the method
is considerably faster: processing of all-sky maps with $\sim
10^{7}$ sky pixels (pixel size of 3'.5) took approximately ten
minutes on a Pentium III 600 MHz computer. This is much less than
required by any existing non-blind separation methods like Maximum
Entropy, taking 6 hours on 30 processors at equivalent resolution
(Stolyarov et al. 2001). Second, we have shown that {\ica} works
in the presence of instrumental noise, although under the
simplifying  assumption that it is white and uniformly
distributed. Third, the algorithm was successfully applied both to
all-sky maps and to limited portions of the sky. Using the full
all sky maps obviously minimizes the sample variance. On the other
hand, we have shown that the foreground removal improves if we cut
regions (around the Galactic plane) where their distribution is
highly non-uniform.
%

Furthermore, we extended the formalism of the separation algorithm
in order to obtain the normalization constants to convert {\ica}
output maps in physical units. We found that the accuracy on this
calibration is limited only by the instrumental noise. The same
noise effect limits the frequency scaling reconstruction. On the
other hand, {\ica} still requires that all the input maps have the
same angular resolution.

We applied {\ica} to different combinations of {\sc Planck}
frequency channels, namely 30, 44, 100, 143, 217, 353 GHz,
containing a mixture of CMB, synchrotron and thermal emissions,
the latter comprising free-free plus thermal dust emission. The
nominal average instrumental noise levels and angular resolutions
of {\sc Planck} have been adopted.

With all these ingredients, {\ica} is able to obtain an excellent
reconstruction (well calibrated) of all these components. The
angular power spectra of both CMB and thermal emissions can be
accurately  reconstructed up to multipoles $\ell\simeq 2000$.
Synchrotron is also recovered on all scales where our template has
significant power, i.e. up to $\ell\simeq 300$. Frequency scalings
and normalization are recovered up to better than 1\% precision.

Nevertheless, a lot of work to assess the real capabilities of
this and other component separation algorithms for CMB experiments
has still to be done. Here we list our main simplifying
assumptions, which we have to relax in future.

The spectral properties of synchrotron and dust emissions have
been assumed to be constant over the sky, although it is known
that both the synchrotron spectral index and the dust temperature
vary from place to place. The spatial pattern of free-free
emission was assumed to follow that of dust, which is certainly
over-simplistic. Spinning dust emission was not taken into
account. We also neglected extra-galactic sources, which are
expected to contribute significantly on all {\sc Planck} channels,
as well as the thermal and kinetic Sunyaev-Zeldovich effect. Very
promising point-source extraction techniques are being extensively
studied (Cay{\'o}n et al. 2000; Vielva et al. 2001).


Finally, the instrumental noise is not uniform across the sky, as
assumed here, but has a complex pattern determined by the adopted
scanning strategy adopted. Also the instrumental beam is not
perfectly circularly symmetric, etc.

Despite all these still unaddressed issues, we demonstrated that
{\ica} is a very promising technique because of three main
features: $(i)$ the good separation achieved on our simulated
maps, which contained noise at the {\sc Planck} nominal level,
$(ii)$ the high computational speed, and $(iii)$ the extreme
flexibility of the code, which is able to work on all-sky maps as
well as on arbitrarily shaped sky regions in a straightforward
way. We recall also that {\ica} is a blind separation algorithm,
able to estimate not only independent components present in the
data, but also their frequency scalings.


\section*{Acknowledgements}

We wish to warmly thank all the {\sc Planck} Low Frequency
Instrument Data Processing Center for having supported this work. 
We gratefully acknowledge financial support from MURST and ASI.

\end{document}